\documentclass[journal=jctcce,manuscript=article]{achemso}
\usepackage[version=3]{mhchem} 

\usepackage{color}
\usepackage{graphicx}
\usepackage{dcolumn}
\usepackage{bm}
\usepackage{multirow}
\usepackage{ulem} 
\usepackage{amsmath}
\usepackage{amssymb}
\usepackage{nameref}

\newcommand{\cpgroup}{{CP~group}}
\newcommand{\fref}[1]{Figure~\ref{#1}} 
\newcommand{\tref}[1]{Table~\ref{#1}} 
\newcommand{\eref}[1]{Eqn.~\ref{#1}} 

\author{Ritama Kar}
\affiliation{ 
Department of Chemistry, Indian Institute of Technology Kanpur (IITK), Kanpur - 208016, India
}%
\author{Sagarmoy Mandal}

\affiliation{{Interdisciplinary Center for Molecular Materials and Computer Chemistry Center, Friedrich-Alexander-Universität Erlangen-Nürnberg (FAU), Nägelsbachstr. 25, 91052 Erlangen, Germany}}
\alsoaffiliation{Erlangen National High Performance Computing Center (NHR@FAU),
Friedrich-Alexander-Universit{\"a}t Erlangen-N{\"u}rnberg, Martensstr. 1, 91058
Erlangen, Germany}
\altaffiliation{Current address: Department of Chemistry, Purdue University, West Lafayette, Indiana 47907, USA}
\author{Vaishali Thakkur}
\affiliation{ 
Department of Chemistry, Indian Institute of Technology Kanpur (IITK), Kanpur - 208016, India
}%
\author{Bernd Meyer}

\affiliation{{Interdisciplinary Center for Molecular Materials and Computer Chemistry Center, Friedrich-Alexander-Universität Erlangen-Nürnberg (FAU), Nägelsbachstr. 25, 91052 Erlangen, Germany}}
\alsoaffiliation{Erlangen National High Performance Computing Center (NHR@FAU),
Friedrich-Alexander-Universit{\"a}t Erlangen-N{\"u}rnberg, Martensstr. 1, 91058
Erlangen, Germany}
\author{Nisanth N. Nair}
\email{nnair@iitk.ac.in}
\affiliation{ 
Department of Chemistry, Indian Institute of Technology Kanpur  (IITK), Kanpur - 208016, India
}%

\title[An \textsf{achemso} demo]
  {Speeding-up Hybrid Functional based {\it {Ab Initio}} Molecular Dynamics using Multiple Time-stepping and Resonance Free Thermostat}

\abbreviations{IR,NMR,UV}

\begin{document}


\begin{abstract}
%
%
%
%
%
{\it Ab initio} molecular dynamics (AIMD) based on density functional theory (DFT) has become a workhorse for studying the structure, dynamics, and reactions in condensed matter systems. Currently, AIMD simulations are primarily carried out at the level of generalized gradient approximation (GGA), which is at the 2$^{\rm nd}$ rung of DFT-functionals in terms of accuracy.
Hybrid DFT functionals which form the 4$^{\rm th}$ rung in the accuracy ladder, are not commonly used in AIMD simulations as the computational cost involved is $100$ times or higher.
To facilitate AIMD simulations with hybrid functionals, we propose here an approach that could speed up the calculations by $\sim~30$ times or more for systems with a few hundred of atoms.
We demonstrate that, by achieving this significant speed up and making the compute time of hybrid functional based AIMD simulations at par with that of GGA functionals, we are able to study several complex condensed matter systems and model chemical reactions in solution with hybrid functionals that were earlier unthinkable to be performed.
\end{abstract}


\section{\label{sec:intro}Introduction}
%
%
%
%
Kohn-Sham density functional theory (KS-DFT) based {\it ab initio} molecular dynamics (AIMD) has been established as a standard tool to investigate structural and dynamic properties of fluids, solids, interfaces, and biological systems.\cite{Chemist's_Guide,Tuckerman_2002_aimd,tuckerman_aimd,marx-hutter-book} 
%
The predictive capability of DFT calculations is largely reliant on the underlying exchange-correlation (XC) functionals.\cite{Chemist's_Guide,Science_DFT_limitations,PRB_SIC,PRL_DFT_errors}
The first and second generations of XC functionals with local density approximation (LDA) and generalized gradient approximation (GGA)\cite{PRA_GGA_Becke,PRB_GGA_LYP,PRL_GGA_PBE} have significant limitations in predicting properties of open-shell systems, band gaps of solids, chemical reaction barriers, and dissociation energies.\cite{Science_DFT_limitations,PRL_DFT_errors}
The major source of inaccuracy in such computations comes from the erroneous inclusion of the unphysical self-interaction of the electron density, caused by residual Coulomb interaction that cannot be eliminated by the exchange part of these functionals. \cite{Chemist's_Guide,Science_DFT_limitations,PRB_SIC,PRL_DFT_errors,JPCL_SIE}
Hybrid functionals can overcome these issues to a large extent, by the inclusion of certain fraction of nonlocal Hartree-Fock (HF) exchange with the standard GGA exchange energy.\cite{Chemist's_Guide,Martin-book,JCP_B3LYP,JCP_PBE0,JCP_HSE}  

Hybrid functional-based AIMD (H-AIMD) simulations are known to improve the prediction of free energetics and mechanisms of chemical processes in liquids and heterogeneous interfaces.\cite{JPCB_AIMD_HFX,JCTC_AIMD_HFX,JCP_AIMD_HFX,Mol_Phy_Car_MLWF,Mol_Phy_Car_MLWF_1,JPCB_water_hfx,Scuseria:JCP:2008,Adamo:2012,JCP_B3LYP,JCP_PBE0,JCP_PBE0_model,JCP_HSE,PCCP_HFX,PCCP_HSE,Galli_RSB_JPCL,Chem_Rev_Cohen,jpcl_2011_hfx,jpcl_2019_hfx,jpcl_2020_hfx}
Yet, the H-AIMD simulations are rarely performed due to the huge computational cost involved.
For AIMD simulations of condensed matter systems, the basis set of choice is plane waves (PWs) as they are complete, orthonormal, inherently periodic, and devoid of basis set superposition error and Pulay forces.\cite{Pople2003,JCP_HFX_Voth}
With PW basis set, H-AIMD simulations are about two orders of magnitude slower than GGA based AIMD simulations for moderate system sizes of $\sim$100 atoms.
If $N_{\rm orb}$ and $N_{\rm G}$ are the number of occupied KS orbitals and the number of PWs, respectively, the HF exchange energy computation scales as $N_{\rm orb}^2 N_{\rm G}\log  N_{\rm G}$.
Typically, the number of PWs is  an order of magnitude larger than the number of atom-centered basis functions, making PW-based hybrid DFT calculations extremely time-consuming. \cite{JCP_HFX_Voth}
It is common to perform a million force evaluations in AIMD simulations, and thus one needs to compute HF exchange integrals a million times during H-AIMD simulations. 
%
Consequently, H-AIMD simulations are rarely performed for more than $\sim 100$ atoms, even though it is more accurate than AIMD with GGA/meta-GGA functionals. 
%

%

%
In order to improve the predictive power of AIMD simulations of complex chemical systems, it is important to perform H-AIMD simulations. 
So far, several efforts have been proposed to speed up such calculations like utilization of localized orbitals,\cite{MLWF_Car,PRB_Car_Wannier,JCP_AIMD_HFX,Mol_Phy_Car_MLWF,Mol_Phy_Car_MLWF_1,Nature_Car_MLWF,PRL_RSB,JCTC_RSB,JCTC_RSB_1,Galli_RSB_CPL,Galli_RSB_JPCL,Galli_RSB_JPCL1,JCP_sagar,Car_hfx_2019,enabling_part2,SeA_RDJ} 
multiple time step (MTS) algorithms,\cite{HFX_Hutter_JCP,MTS_AIMD_Ursula,MTS_AIMD_Steele_3} 
{coordinate-scaling,}\cite{JPCL_2018_Bircher,CPC_BIRCHER_2020}
 massive parallellization\cite{HFX_Curioni,DUCHEMIN_2010,VARINI_2013,BARNES_2017,sagar_JCC_scaling}
and others.\cite{jctc_Bolnykh_2019,single_precision_hfx,HFX_JB,HFX_Goedecker}

Recently, a strategy named multiple time stepping with adaptively compressed exchange (MTACE) has been proposed in our group, which uses the adaptively compressed exchange (ACE) operator formulation\cite{ACE_Lin,ACE_Lin_1,ACE_2023} and MTS scheme to significantly reduce the computational cost of these simulations.\cite{JCP_2019_sagar,sagar_JCC} 
%
%
%
In the MTACE method,
a modified self-consistent field (SCF) iteration procedure is used where the ACE operator constructed at the first SCF step 
is used for the subsequent SCF steps to compute exchange energy.
The construction of the ACE operator is as time consuming as that for the exact exchange operator, while applying the ACE operator to compute the exchange energy has a negligible computational cost.
%
The ionic force computed by applying ACE operator constructed at the first SCF step is denoted as ${\textbf F}^{\textrm{ACE}}$, which is different from the exact ionic force ${\textbf F}^{\textrm{exact}}$ computed with explicit construction of the exact exchange operator at each step of SCF iteration.
%
%
%
%
Now, we write
\begin{equation}
  {\textbf F}^{\textrm{exact}} = {\textbf F}^{\rm fast} + {\textbf F}^{\rm slow} \enspace ,
\end{equation}
where, ${\textbf F}^{\rm fast}$ and 
${\textbf F}^{\rm slow}$ are fast and slow varying force components, respectively.
We assign these components as 
\begin{eqnarray}
\label{ffast}
{\textbf F}^{\textrm{fast}} \equiv {\textbf F}^{\textrm{ACE}} \enspace , \label{e:f:fast}
\end{eqnarray}
and
\begin{eqnarray}
\label{fslow}
{\textbf F}^{\textrm{slow}} \equiv \Delta {\textbf F} = \left ( {\textbf F}^{\rm exact} -{\textbf F}^{\textrm{ACE}} \right ) \enspace . \label{e:f:slow}
\end{eqnarray}
It was  observed that $\Delta \mathbf F$ is a slowly varying component of force and such a splitting of force is a reasonable assumption.\cite{JCP_2019_sagar} %
It is worth noting that computation of ${\textbf F}^{\textrm{ACE}}$ is cheap, while $\Delta \textbf F$ is expensive to compute.

With these assumptions, the Liouville operator (${\mathcal L}$) of a system containing $N_{\rm at}$ atoms, having `slow' and `fast' ionic force components can be expressed as
\begin{equation}
\label{mts-l}
 i{\mathcal L}=i{\mathcal L}_V^{\rm fast}+i{\mathcal L}_V^{\rm slow}+i{\mathcal L}_X \enspace ,
\end{equation} 
with 
\begin{equation}
\label{mts-l-component}
    i{\mathcal L}^{\rm fast}_V  =\sum_{I=1}^{3N_{\rm at}} \frac{F_I^{\rm fast}}{M_I} \frac{\partial}{\partial V_I} , 
    \enspace \enspace \enspace
    i{\mathcal L}^{\rm slow}_V  =\sum_{I=1}^{3N_{\rm at}} \frac{F_I^{\rm slow}}{M_I} \frac{\partial}{\partial V_I}
\end{equation}
and 
\begin{equation}
    i{\mathcal L}_X =\sum_{I=1}^{3N_{\rm at}} V_I\frac{\partial}{\partial X_I} \enspace .
\end{equation}
Here, Cartesian coordinates and the velocity component of any degree of freedom $I$ are denoted by $X_I$ and $V_I \equiv \dot{X}_I$, respectively.
The MTACE method employs the reversible reference system propagator algorithm (r-RESPA),\cite{respa}
where a symmetric Trotter factorization of the classical time evolution operator is expressed as,
%
\begin{equation}
\label{trotter}
 e^{i{\mathcal L}\Delta t}\approx e^{i{\mathcal L}_V^{\rm slow}\Delta t/2}\left[e^{i{\mathcal L}_V^{\rm fast}\delta t/2}e^{i{\mathcal L}_X \delta t}e^{i{\mathcal L}_V^{\rm fast}\delta t/2}\right]^{n_{\rm MTS}} e^{i{\mathcal L}_V^{\rm slow}\Delta t/2} \enspace .
\end{equation}
During the propagation of the system according to the propagator in \eref{trotter},
the computationally costly $\Delta {\textbf F}$ (i.e., ${\textbf F}^{\textrm{slow}}$) force is computed only once at every $n_{\rm MTS}$ step,
while the 
computationally cheap force
${\textbf F}^{\textrm{ACE}}$ (i.e., ${\textbf F}^{\textrm{fast}}$) 
is computed at every MD step $\delta t$ (also called as inner time step).
Here, $n_{\rm MTS} = \Delta t/\delta t$, is the MTS time step factor, where $\Delta t$ is
the larger time step (also called as outer time step) which controls the update using the
$\Delta \mathbf F$ force.
%
%
Thus, the computational efficiency of the MTACE method in generating classical trajectory of atoms depends on $n_{\rm MTS}$.
With increasing $n_{\rm MTS}$, the explicit construction of the exact exchange operator becomes less frequent, thereby reducing the computational cost significantly.
%
%
In an attempt to further improve the MTACE approach, the selected column of the density matrix (SCDM)\cite{SCDM_main} method was used to screen the occupied orbitals based on their spatial overlap during the construction of the ACE operator. 
%
%
%
%
%
%
Such an implementation could yield one order of magnitude speed up for a system of $\sim$100 atoms, without compromising on the accuracy of any computed structural and dynamical properties.\cite{sagar_JCTC} 
Subsequently, to take advantage of large computational resources available on modern supercomputers, we implemented MTACE using a task group based parallelization strategy.\cite{KLOFFEL2021,sagar_JCC_scaling} 
Our benchmark calculations revealed that, with these developments, H-AIMD simulations can be carried out at a much lower computational cost than that was possible before.

In our prior study,\cite{JCP_2019_sagar,sagar_JCTC} it was found that 
there is a limit beyond which $n_{\rm MTS}$ can not be increased, and for typical molecular systems with H atoms, the best $\Delta t$ value was found to be $7.2$~fs, which implies $n_{\rm MTS} =
15$ while using $\delta t = 0.48$~fs.
%
%
The obstacle in increasing $n_{\rm MTS}$ beyond a certain value is the occurrence of resonance,\cite{Resonace_1,Resonace_2,Resonace_3} 
which 
restrict on the largest value that can be chosen for $\Delta t$. 
%
%
%
%
To overcome resonance effects in r-RESPA integration,  resonance free thermostats were proposed.\cite{Resonace_free_1,Resonace_free_2,Resonace_free_3,Resonace_free_4,Resonace_free_5,Resonace_free_6,SINR_main} 
%
%
%
%
In this work, we have used a chain of stochastic Nos{\'e}--Hoover chain (NHC) thermostats via isokinetic constraint named as stochastic isokinetic Nos{\'e}--Hoover (RESPA) or {SIN(R)}.\cite{SINR_main,SINR_polarised,SINR-middle,SINR_MP_21,SINR_20_jctc,SINR_21_epj}
We investigated how this method facilitates the increase in $\Delta t$ ( or $n_{\rm MTS}$), and thereby its impact on the efficiency of the MTACE based H-AIMD simulations.
%
%
%
We call the new approach as resonance free MTACE or in abbreviation RF-MTACE.
%
%
%
%
%
%
First, we assessed the accuracy and computational performance of the proposed method on computing the properties of bulk water using RF-MTACE.
%
%
%
%
Then we benchmarked the computational performance of RF-MTACE for several other interesting systems with varying complexity:
formamide in water, benzoquinone in methanol, $\rm {Fe}^{3+}$ in water, rutile-TiO$_2$ (110) surface with defect, and a protein-ligand complex. 
Finally, we demonstrated a real life application of the method by employing it in computing the free-energy barrier and mechanism of a chemical reaction in solution.

\section{\label{sec:methods}Methods}

\subsection{MTACE Method for H-AIMD Simulation}
\begin{figure}[h]
	\centering
		\includegraphics[scale=0.7]{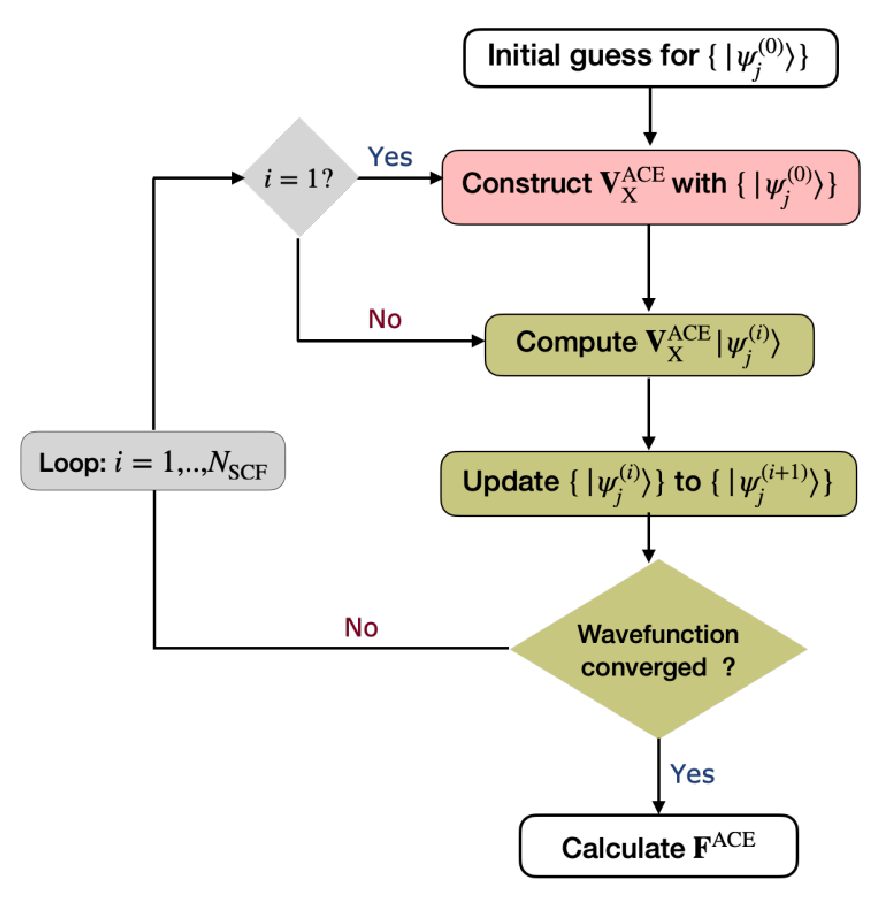}
	    \caption{Flowchart showing the SCF procedure involving exchange operator ${\mathbf V}_{\rm X}^{\rm ACE}$ in the MTACE method. ${\mathbf V}_{\rm X}^{\rm ACE}|\psi _{j}\rangle$ is the gradient on $|\psi _{j}\rangle$ due to the exact exchange part.
        }
	\label{f:SCF-MTACE}
\end{figure}
In a typical KS-DFT calculation with hybrid functionals, the exact exchange operator ${\mathbf V}_{\rm X}$ is defined as
\begin{equation}
{\mathbf V}_{\rm X}= -\sum_{j}^{N_{\rm orb}} \frac{| \psi _{j} \rangle  \langle \psi _{j} |}{r_{12}} \enspace ,
\end{equation}
where $\{|\psi_{j} \rangle\}$ is the set of occupied KS orbitals.
The total number of occupied orbitals is $N_{\rm orb}$ and $r_{12}=\left | \mathbf r_1 - \mathbf r_2 \right | $.
This operator is applied on the KS orbitals $|\psi _{i} \rangle$ as
%
\begin{equation}
\label{vx}
\begin{split}
{\mathbf V}_{\rm X}|\psi _{i}\rangle & =- \sum_{j}^{N_{\rm orb}} |\psi _{j} \rangle \left \langle\psi _{j} \left | \left ( r_{12}\right )^{-1} \right | \psi _{i}\right \rangle   \\ & =- \sum_{j}^{N_{\rm orb}} v_{ij}(\mathbf{r}) |\psi _{j}\rangle \enspace, \enspace ~~i=1,...,N_{\rm orb} \enspace , 
\end{split}
\end{equation}
with
\begin{equation}  
\label{vij}
v_{ij}(\mathbf {r})=\left \langle\psi _{j} \left | \left ( r_{12}\right )^{-1} \right | \psi _{i}\right \rangle \enspace  . 
\end{equation}
HF exact exchange energy is computed as
\begin{equation}
\label{HFX}
E^{\textrm{HF}}_{\textrm{X}} = - \sum_{i,j}^{N_{\rm orb}}
\left \langle \psi _{i} \left | v_{ij} \right | \psi _{j}
\right \rangle \enspace .
\end{equation}

In the ACE operator formalism\cite{ACE_Lin,ACE_Lin_1}, the ACE operator (${\mathbf V}_{\rm X}^{\rm ACE}$) is constructed through a series of linear algebra operations:
\begin{equation}
\label{vx_ace}
{\mathbf V}_{\rm X}^{\rm ACE}   = - \sum_{k}^{N_{\rm orb}} |P_{k} \rangle  \langle P_k |   \enspace,
\end{equation}
with $\{|P_{k} \rangle\}$ as the columns of the matrix ${\mathbf P}$, which is defined 
as
\begin{equation}
  {\mathbf P}= {\mathbf W}{\mathbf L}^{-T} \enspace .
\end{equation}
Here, the columns of the matrix ${\mathbf W}$ can be computed as
\begin{equation}
|W_{i}\rangle={\mathbf V}_{\rm X}|\psi _{i}\rangle \enspace , \enspace ~~i=1,...,N_{\rm orb} \enspace ,
\end{equation}
and ${\mathbf L}$ is a lower triangular matrix resulting from the Cholesky factorization of matrix $-{\mathbf M}$, with elements
\begin{equation}
(\mathbf M)_{kl}= \left \langle \psi_k | {\mathbf V}_{\rm X}|\psi _{l} \right \rangle   \enspace.
\end{equation}

\begin{figure}[h]
    \centering
    \includegraphics[scale=0.5]{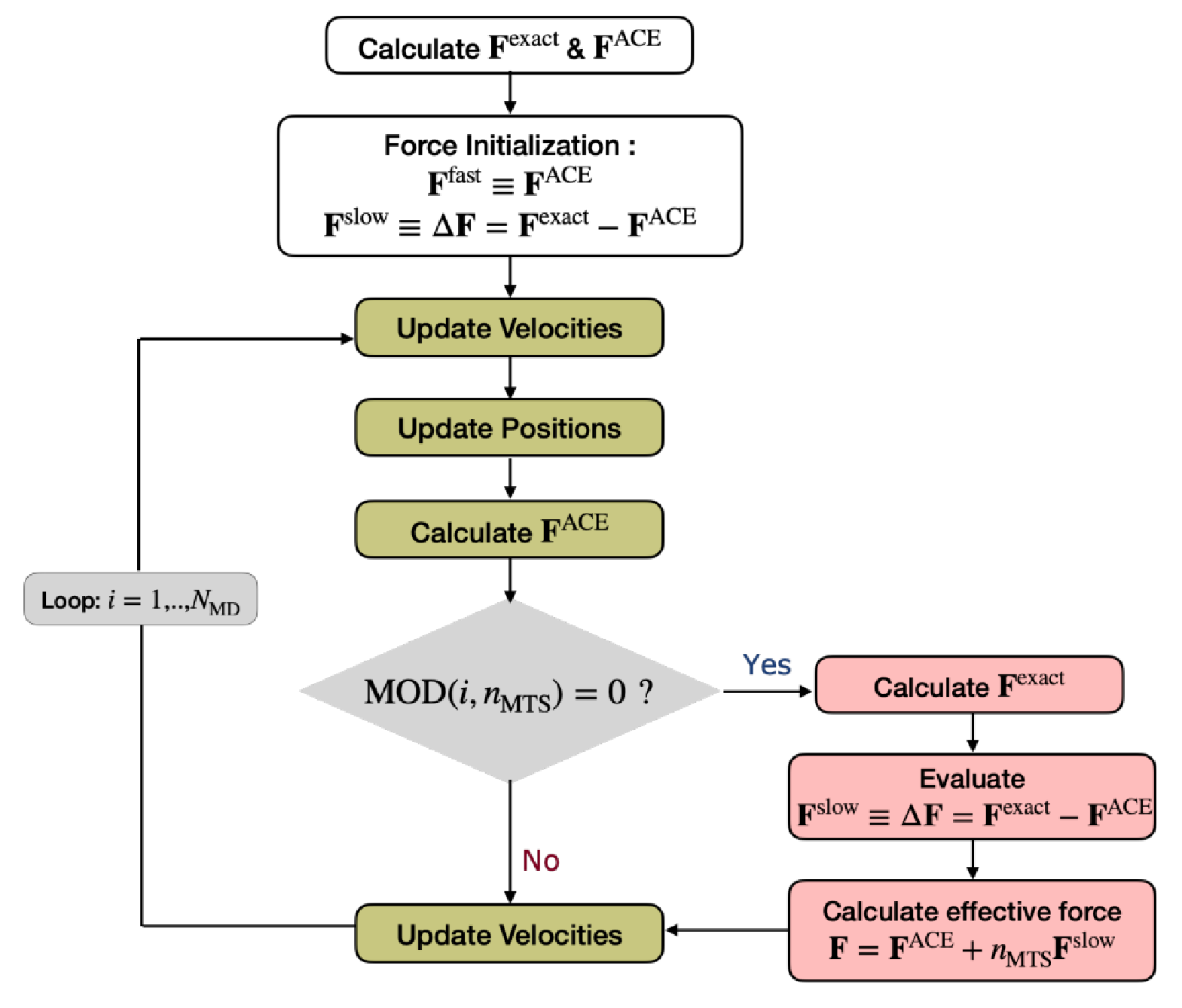}
    \caption{Flowchart illustrating the MTACE method involving the contributions due to exact exchange.}
    \label{fig:mtace}
\end{figure}

As the application of the ${\mathbf V}_{\rm X}^{\textrm{ACE}}$ operator on each KS orbitals is computationally cheaper than 
${\mathbf V}_{\rm X}$ operator, this operator can be employed in different ways to speed up H-AIMD based simulations.\cite{sagar_JCC,JCP_2019_sagar,sagar_JCC_scaling,sagar_JCTC,ACE_2023,SeA_RDJ}
In the MTACE method, a modified SCF procedure is used as shown in \fref{f:SCF-MTACE}. 
%
The ACE operator ${\mathbf V}_{\rm X}^{\textrm{ACE}}$ constructed in the first SCF step is used in the subsequent SCF steps.
On using this approximation, the set of optimized orbitals obtained at the end of the SCF procedure is different to the one obtained using the exact operator ${\mathbf V}_{\rm X}$.
Thus, the force $\mathbf F^{\textrm{ACE}}$ on the atoms computed using the optimized orbitals employing the ACE approximation 
deviates from the exact atomic forces $\mathbf F^{\rm exact}$.
As shown earlier\cite{JCP_2019_sagar,sagar_JCTC}, the difference $\Delta \mathbf F \equiv \mathbf F^{\rm exact} - \mathbf F^{\textrm{ACE}}$ is small and is slowly varying.
Thus $\mathbf F^{\textrm{ACE}}$ and $\Delta \mathbf F$ can be considered as fast force and slow force, respectively, as in \eref{ffast} and \eref{fslow} and can be treated using the r-RESPA integration scheme.
The algorithm for the MTACE method is given in \fref{fig:mtace}.

As discussed in the earlier work,\cite{sagar_JCTC}
one can construct exchange operators using localized orbitals
$\{|\phi_{j} \rangle\}$
obtained using the SCDM approach,\cite{SCDM_main} which is called s-MTACE method hereafter.
%
%
%
In this method, a cut-off 
\[ \int d\textbf{r} \left | \phi_{i}(\textbf{r}) \phi_{j}^{*}(\textbf{r}) \right | \geqslant \rho_{\textrm {cut}} \enspace,\]
is introduced to screen the orbital pairs.
 Orbital pairs $i$-$j$ fulfilling the above criteria is only considered in the computation of  $\mathbf V_{\rm X}^{\rm ACE}$.
Such a screening reduces the number of orbitals accounted in the exact exchange computations.\cite{SCDM_main}

\subsection{\label{sec:cp-group}Task Group Approach}

%
%
%
%
\begin{figure}[h]
    \centering
    \includegraphics[scale=0.42]{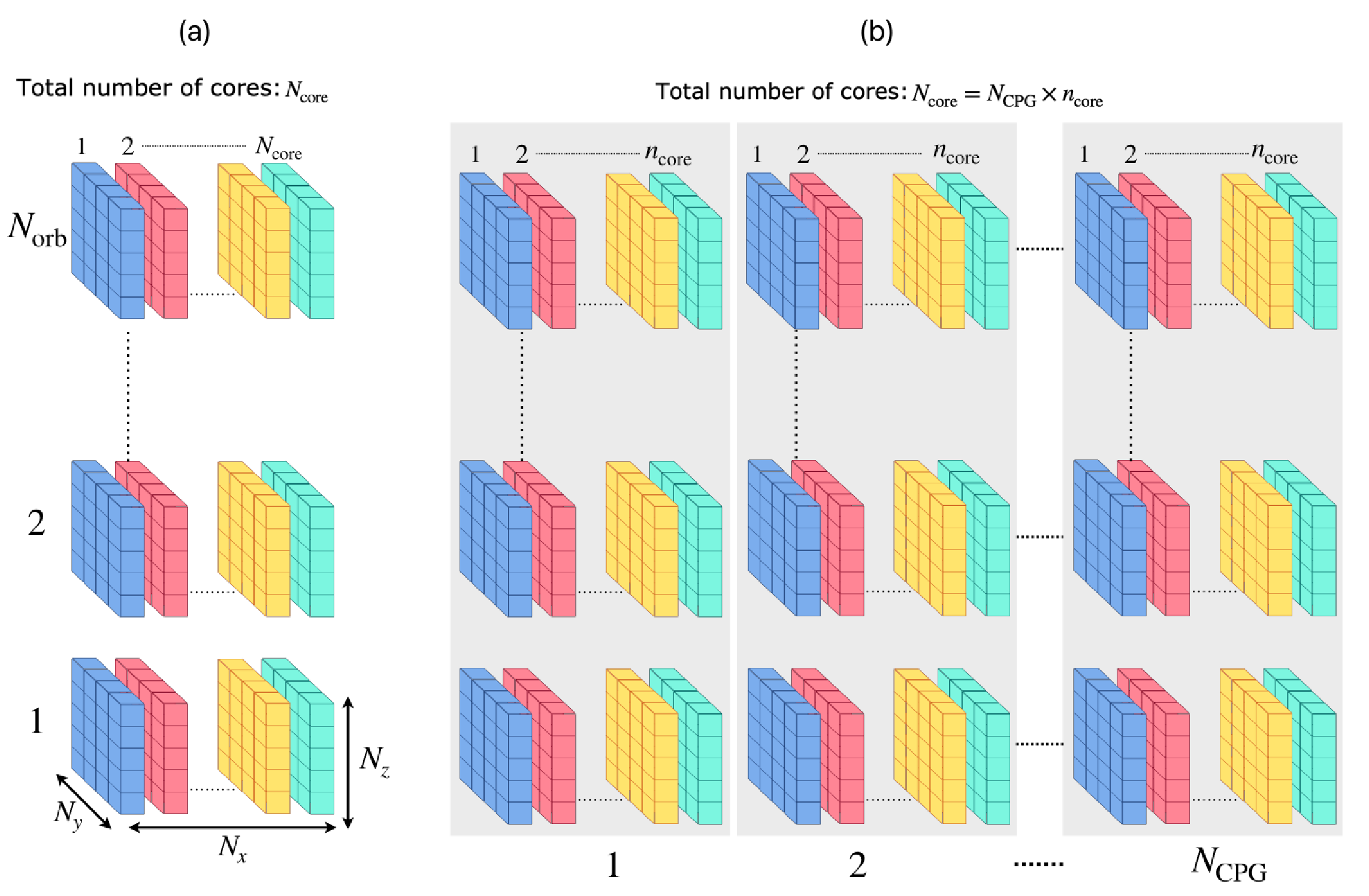}
    \caption{(a) Conventional distribution of wavefunctions on grids into $N_{\rm core}$ compute cores. 
    Total number of grids in real space: $N_{\rm R}=N_xN_yN_z$.
    Each computr core stores $N_x/N_{\rm core}$ $yz$ planes of wavefunctions.
    (b) {\cpgroup} data distribution is shown for $N_{\rm core}$ compute cores, which are divided into $N_{\rm CPG}$ CP groups. Each {\cpgroup} contains $n_{\rm core}$ compute cores.}
    \label{fig:cpgroup}
\end{figure}
In typical implementation of PW-DFT within the CPMD code\cite{cpmd}, the three-dimensional fast Fourier transforms (FFT) grids are divided  into slabs along $X$ direction and the slabs are  distributed among the compute cores.
In slab decomposition, 
each compute core stores  $N_{ x}/N_{\rm core}$ $yz$ planes of wavefunctions, where $N_{x}$ is the number of FFT grid points in the $X$-direction and $N_{\rm core}$ is the number of compute cores; see \fref{fig:cpgroup}(a).
%
%
The performance of such an implementation in terms of strong scaling is limited by $N_{x}$.
In a standard application of PW-DFT, $N_{x}$ is a few hundred, whereas the number of compute cores accessible on any modern supercomputing resource is a few thousand to millions. 
As a result, resource utilization is not efficient with slab decomposition. 
In order to address this issue, a parallelization technique called {\cpgroup}, based on MPI task groups, was proposed.\cite{HFX_Curioni,CPG_curioni}
%
%
%
%
%
The {\cpgroup} methodology involves partitioning the total number of compute cores, into $N_{\rm CPG}$ distinct groups. 
Each group is assigned $n_{\rm core}$ compute cores, where $n_{\rm core} = N_{\rm core} / N_{\rm CPG}$,
and keeps a complete copy of the wavefunctions, which is then distributed across the $n_{\rm core}$ compute cores within the group. 
%
%
%
%
%
Consequently, every compute core of a group holds $N_{x}/n_{\rm core}$ $yz$ planes of wavefunctions, as shown in \fref{fig:cpgroup}(b).
%
The distribution of workload across the task groups involves parallelization of computations over orbital pairs contributing to the exchange integral (\eref{HFX}). 
This parallelization strategy ensures that calculations within each task group are restricted to a specific subset of orbital pairs.
%
%
Using this approach, the calculation of $v_{ij}(\mathbf{r})$ (\eref{vij}) for all orbital pairs can be efficiently executed in segments within the designated group.
%
Finally, a global summation is performed across all groups in order to obtain the full exchange operator.
%
%
%
%
This strategy also
minimizes inter-group communications.
%
In situations where $N_{\rm core}>>N_{x}$, it has been demonstrated that the {\cpgroup} approach can yield excellent scaling performance for hybrid DFT calculations.\cite{HFX_Curioni,CPG_curioni,KLOFFEL2021,sagar_JCC_scaling}

%

\subsection{\label{sec:sinr}SIN(R) Thermostat}
To address the resonance problems arising in RESPA based integration, Leimkuhler {\it et al.} proposed a new class of thermostats with iso-kinetic constraints.\cite{Isokinetic_1,Evans-Morriss-book,Isokinetic_3}
In their approach, a chain of thermostat variables is coupled to every physical degrees of freedom and the total kinetic energy is preserved. 
Later, a stochastic modification was introduced to improve the efficiency of the thermostat.\cite{SINR_main}
%
%
%
In this article, we are briefly presenting the SIN(R) thermostat for completion of discussions, although detailed discussions can be found elsewhere.\cite{SINR_main,SINR_polarised,SINR-middle,SINR_MP_21,SINR_20_jctc,SINR_21_epj}
%
%
%
%
The equations of motion are described as follows:
%
\begin{equation}
  \label{eom}  
    \begin{split}
        dX_I &= V_I dt \enspace , \\
        dV_I &= \left[\frac{F_I\mathbf{(X)}}{M_I} -\lambda_I V_I\right]dt \enspace , \\
        dv_{1,I,j} &=-\lambda_I v_{1,I,j}dt - v_{2,I,j}v_{1,I,j}dt \enspace , \\
        dv_{2,I,j} &=\left[\frac{G(v_{1,I,j})}{Q_2}-\gamma v_{2,i,j}\right]dt+\sigma dW_I \enspace ,
    \end{split}    
\end{equation}
where $I=1,\cdots,3N_{\rm at}$ runs over all the physical degrees of freedom, and $j=1,\cdots,L$, with $L$ representing the length of the thermostat chain.
%
$v_{1,I,j}$ and $v_{2,I,j}$ are velocities of the thermostat auxiliary variables.
%
Here, $\gamma$, $T$ and $k_{\rm B}$ denote the frictional constant, temperature and Boltzmann's constant, respectively, while $\sigma=\sqrt{{2k_{\rm B}T\gamma}/{Q_2}}$.
%
$dW_I$ defines small increment of Weiner process, and
%
\begin{equation}
\label{geq}
    G(v_{1,I,j})=Q_1 v_{1,I,j}^2-k_{\rm B} T \enspace .
\end{equation}
%
%
The thermostat mass parameters $\{Q_i\}$ are related to a time constant $\tau$ as 
$Q_i = k_{\rm B} T \tau ^2$.
In this case, iso-kinetic constraint on every degree of freedom $I$ is expressed as
%
 \begin{equation}
 \label{kecons}
    \Lambda_I = M_I V_I^2 + \frac{L}{L+1}\sum_{j=1}^L Q_1 v_{1,I,j}^2 = Lk_{\rm B} T  \enspace ,
 \end{equation}
with Lagrange multiplier $\lambda_I$ given by 
\begin{equation}
    \label{lag}
    \lambda_I=\frac{1}{\Lambda_I} \left [ V_I F_I-\frac{L}{L+1}\sum_{j=1}^L Q_1 v_{1,I,j}^2 \, v_{2,I,j} \right ] \enspace .
\end{equation}

To make use of r-RESPA, we write\cite{respa,SINR_main}
\begin{equation}
    \label{lag-r}
    \lambda_I = \lambda_I^{\rm fast} + \lambda_I^{\rm slow} + \lambda_I^{{\rm N}} \enspace ,
\end{equation}
where
\begin{equation}
    \begin{split}
        \lambda_I^{\rm fast} & = \frac{1}{\Lambda_I}  V_I F_I^{\rm fast} \enspace ,\\
        \lambda_I^{\rm slow} & = \frac{1}{\Lambda_I} V_I F_I^{\rm slow} \enspace ,\\
        \lambda_I^{{\rm N}} & = -\frac{1}{{\Lambda_I}} \frac{L}{L+1}\sum_{j=1}^L Q_1 v_{1,I,j}^2 \, v_{2,I,j} \enspace . 
    \end{split}
\end{equation}
The Liouville operator is split as ~\cite{SINR_main, SINR_polarised, SINR-middle} 
\begin{equation}
\label{sinr-l}
 i{\mathcal L}=i\Tilde{\mathcal L}_V^{\rm fast}+i\Tilde{\mathcal L}_V^{\rm slow}+i{\mathcal L}_X+i{\mathcal L}_{\rm N}+i{\mathcal L}_{\rm OU} \enspace ,
\end{equation}
where 
\begin{equation}
\label{sinr}
\begin{split}
i{\mathcal L}_X &=\sum_{I=1}^{3N_{\rm at}} V_I\frac{\partial}{\partial X_I} \enspace ,\\
i\Tilde{\mathcal L}^{\rm fast}_V & =\sum_{I=1}^{3N_{\rm at}} \left[\left(\frac{F_I^{\rm fast}}{M_I}-\lambda_I^{\rm fast} V_I\right) \frac{\partial}{\partial V_I} - \lambda_{I}^{\rm fast}\sum_{j=1}^L v_{1,I,j} \, \frac{\partial}{\partial v_{1,I,j}} \right] \enspace ,\\
i\Tilde{\mathcal L}^{\rm slow}_V & =\sum_{I=1}^{3N_{\rm at}} \left[\left(\frac{F_I^{\rm slow}}{M_I}-\lambda_I^{\rm slow} V_I\right) \frac{\partial}{\partial V_I} - \lambda_{I}^{\rm slow}\sum_{j=1}^L v_{1,I,j} \, \frac{\partial}{\partial v_{1,I,j}} \right] \enspace ,\\
i{\mathcal L}_{\rm N} & = -\sum_{I=1}^{3N_{\rm at}} \left[\lambda_I^{\rm N} V_I \frac{\partial}{\partial V_I} + \lambda_I^{\rm N} \sum_{j=1}^L v_{1,I,j} \frac{\partial}{\partial v_{1,I,j}}+ \sum_{j=1}^L v_{2,I,j} \, v_{1,I,j} \, \frac{\partial}{\partial v_{1,i,j}} - \sum_{j=1}^L \frac{G(v_{1,I,j})}{Q_2} \, \frac{\partial}{\partial v_{2,I,j}} \right] \enspace ,\\
i \mathcal L_{\rm OU} & =\sum_{I=1}^{3N_{\rm at}} \sum_{j=1}^L \left [-\frac{\partial}{\partial v_{2,I,j}}(\gamma v_{2,I,j})+\frac{\gamma k_{\rm B}T}{Q_2} \frac{\partial ^2}{\partial v_{2,I,j}^2} \right ] \enspace .
\end{split}
\end{equation}
%
Using these expressions, we modify \eref{trotter}
while employing the r-RESPA method during the propagation of the system.
%
\subsection{\label{sec:rfmtace}RF-MTACE Method for H-AIMD Simulations}
\begin{figure}[h]
	\centering
		\includegraphics[scale=0.5]{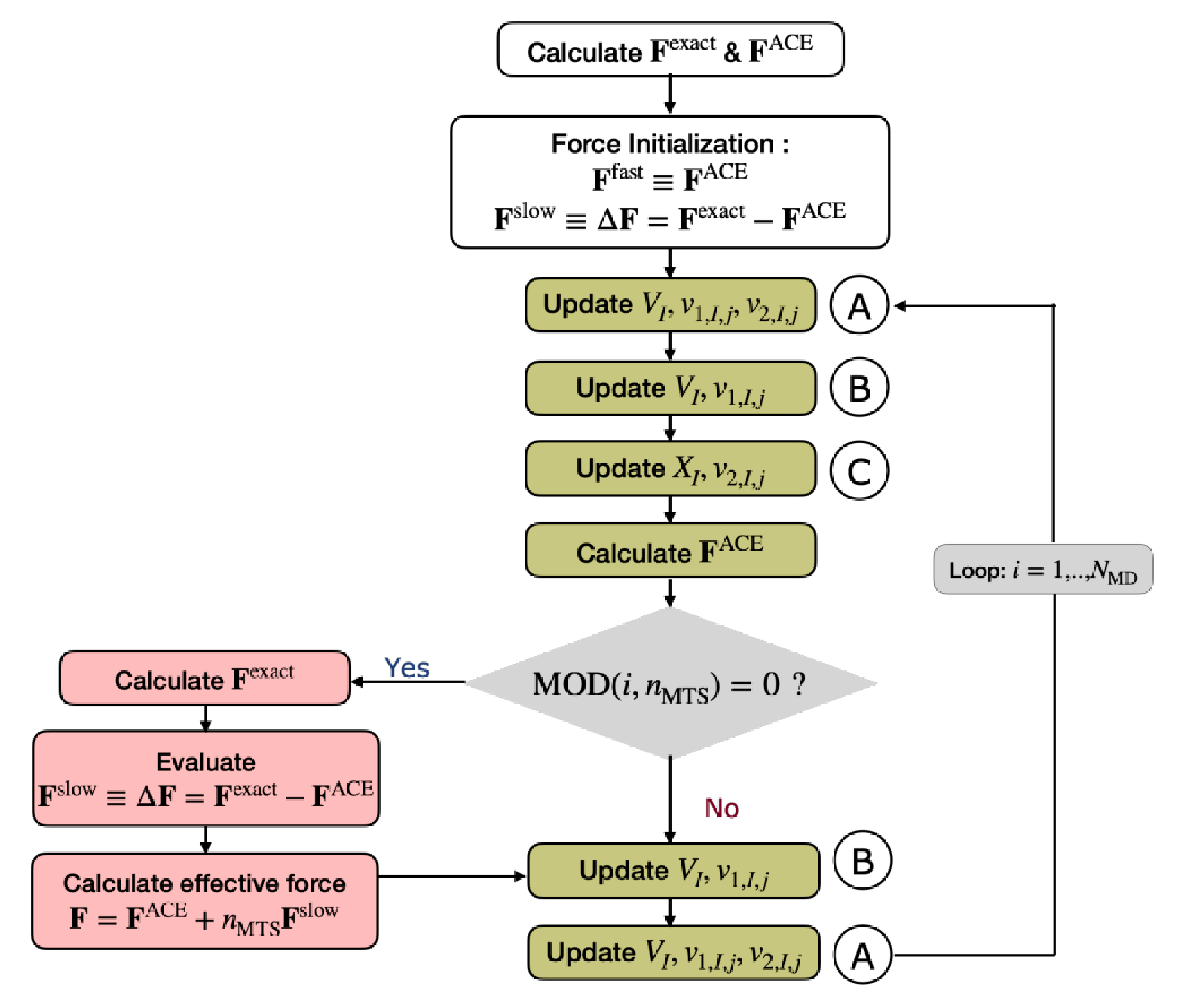}
	    \caption{Flowchart showing the algorithm of H-AIMD simulation using the RF-MTACE method. The equations for the updates, labeled as A, B and C, are mentioned in the \nameref{appendix}.}
	\label{fig:rfmtace}
\end{figure}
Several approaches have been proposed to factorize the time evolution operator in \eref{sinr-l}.\cite{SINR-middle,SINR_polarised,SINR_MP_21,SINR_20_jctc,SINR_21_epj}
Here, we use 
the eXtended-Inner-RESPA or XI-RESPA \cite{SINR_main} to integrate the equations of motion as given below:
\begin{equation}
\label{trotter-sinr}
\begin{split}
 e^{i{\mathcal L}\Delta t} &\approx \left[e^{i{\mathcal L}_{\rm N}\delta t/2}e^{(i\Tilde{\mathcal L}_V^{\rm fast}+n_{\rm MTS}i\Tilde{\mathcal L}_V^{\rm slow})\delta t/2}e^{i{\mathcal L}_X\delta t/2}e^{i{\mathcal L}_{\rm OU}\delta t} e^{i{\mathcal L}_X\delta t/2}e^{i\Tilde{\mathcal L}_V^{\rm fast}\delta t/2} e^{i{\mathcal L}_{\rm N}\delta t/2}\right] \\
 & \times \left[e^{i{\mathcal L}_{\rm N}\delta t/2}e^{i\Tilde{\mathcal L}_V^{\rm fast}\delta t/2}e^{i{\mathcal L}_X\delta t/2}e^{i{\mathcal L}_{\rm OU}\delta t}e^{i{\mathcal L}_X\delta t/2}e^{i\Tilde{\mathcal L}_V^{\rm fast}\delta t/2}e^{i{\mathcal L}_{\rm N}\delta t/2}\right]^{n_{\rm MTS}-2} \\
 & \times \left[e^{i{\mathcal L}_{\rm N}\delta t/2}e^{i\Tilde{\mathcal L}_V^{\rm fast}\delta t/2}e^{i{\mathcal L}_X\delta t/2}e^{i{\mathcal L}_{\rm OU}\delta t}e^{i{\mathcal L}_X\delta t/2}e^{(i\Tilde{\mathcal L}_V^{\rm fast}+n_{\rm MTS}i\Tilde{\mathcal L}_V^{\rm slow})\delta t/2}e^{i{\mathcal L}_{\rm N}\delta t/2} \right] \enspace .
 \end{split}
\end{equation}
%
%
%
The above propagator can be implemented as illustrated in \fref{fig:rfmtace}.
The equations for updating the SIN(R) variables, velocities, and positions are provided in the Appendix.
%
%
%
%
\section{Computational Details}
A series of hybrid functional based Born-Oppenheimer MD (BOMD) simulations were performed in the NVT ensemble at 300K using a modified version of the CPMD code\cite{cpmd,KLOFFEL2021}, where RF-MTACE is implemented.
%
%
We employed the norm-conserving Troullier-Martin type pseudopotentials\cite{PRB_TM} generated with GGA functional to describe the core electrons. 
A energy cutoff of 80~Ry was used to expand the wavefunctions in PW basis set.
At every MD step, we converged the wavefunctions using a direct minimization approach till the magnitude of the  wavefunction gradient was below $1\times 10^{-6}$ au.
We employed either direct inversion of the iterative subspace (DIIS)\cite{PULAY_DIIS,HUTTER_DIIS} or preconditioned conjugate gradient (PCG)\cite{PRB_PCG} method for minimization of wavefunctions.
Always stable predictor corrector extrapolation scheme\cite{JCC_ASPC} of order 5 was used to obtain initial guess of wavefunctions.
To benchmark the performance, four sets of calculations were conducted: {\bf GGA}, {\bf VV}, {\bf MTACE}-$n_{\rm MTS}$ and {\bf RF-MTACE-$n_{\rm MTS}$}; see \tref{table:sim} for details.
{\bf MTACE}-$n_{\rm MTS}$
and
{\bf RF-MTACE}-$n_{\rm MTS}$
runs were performed using the s-MTACE approach.
For the {\bf MTACE}-$n_{\rm MTS}$ runs, $n_{\rm MTS} = 15$ was chosen.
\begin{table}[t]
    \centering
    \begin{tabular}{|p{4cm}|p{3cm}| p{3cm}| p{4cm}|} \hline \hline
       { Simulation label}      & Functional & BOMD scheme & Thermostat  \\ \hline
       {\bf GGA}     & GGA & Conventional & Massive NHC  \\ \hline
       {\bf VV}      & Hybrid & Conventional & Massive NHC   \\ \hline
       {\bf MTACE}-$n_{\rm MTS}$ & Hybrid & s-MTACE & Massive NHC \\ \hline
       {\bf RF-MTACE}-$n_{\rm MTS}$ & Hybrid & s-MTACE & SIN(R) \\ \hline \hline
    \end{tabular}
    \caption{Definition of labels for different simulations carried out in this work.}
    \label{table:sim}
\end{table}
    %
    In {\bf RF-MTACE}-$n_{\rm MTS}$
runs, we used the SIN(R) thermostat with the following parameters: 
    $L=4$, $\tau=9.7$~fs, and $\gamma=0.01$~fs$^{-1}$. 
    %
    %

     Benchmark calculations were performed on six different systems: bulk water, formamide in basic solution,  benzoquinone radical anion in methanol, Fe$^{3+}$ in water, TiO$_2$ surface with oxygen vacancy, and drug bound class-C $\beta$-Lactamase (CBL) enzyme.
     %
     %
     Additionally, we used the {\bf RF-MTACE}-$n_{\rm MTS}$ method for computing the free energy surface of formamide hydrolysis in basic solution with the aid of an enhanced sampling technique. 
    Detailed descriptions of the systems are provided below, and 
    some of the crucial system and simulation parameters 
    are listed in \tref{tab:systems}.
    For all the runs, we used $\delta t =0.48$~fs, except in the case of TiO$_2$ surface, where $\delta t =0.96$~fs was taken.
    We used PBE (GGA) and PBE0 (hybrid) functionals unless otherwise mentioned.
    For the systems with a net charge, a homogeneous counter background charge was added to maintain the charge neutrality.
    %
    
    %
    %
%

%

\begin{table}[]
\centering
\caption{\label{tab:systems}
System and simulation parameters:
%
%
%
$n_{\rm MTS}$ is the MTS time step factor used in {\bf RF-MTACE}-$n_{\rm MTS}$ simulations.
$\delta t$ is the MD time step used in {\bf GGA} and {\bf VV} calculations.
The same time step was used as inner time step in {\bf RF-MTACE}-$n_{\rm MTS}$ runs.
$2S+1$ denotes the total spin multiplicity.
The cutoff for the screening of SCDM-localized orbitals is denoted by $\rho_{\rm cut}$. 
}
\begin{tabular}{|l|c|c|c|c|c|c|c|c|c|}
\hline \hline
System & Charge & $2S+1$ & $N_{\rm at}$ & $n_{\rm core}$ & $N_{\rm CPG}$ & $n_{\rm MTS}$ & $\delta t$(fs) & $\rho_{\rm cut}$  \\
\hline
32 water & 0 & 1 & 96 & 120 & 1 & 250$^{[1]}$ & 0.48 & $2.5\times 10^{-2}$ \\
128 water & 0 & 1 & 384 & 180 & 1 & 200 & 0.48 & $2.0\times 10^{-3}$ \\
%
Formamide solution & -1 & 1 & 95 & 120 & 1 & 200 & 0.48 & $2.5\times 10^{-2}$ \\
BQ$^{.-}$ in MeOH & -1 & 2 & 354 & 192 & 1 & 200 & 0.48 & $2.0\times 10^{-3}$ \\ 
Fe$^{3+}$ in water & $+3$ & 6 & 193 & 144 & 1 & 200 & 0.48 & $1.0\times 10^{-2}$ \\
TiO$_{2-x}$ surface & 0 & 3 & 143 & 240 & 4 & 200 & 0.96 & $2.0\times 10^{-3}$ \\
Enzyme:Drug & +1 & 1 & $76^{[2]}$ & 240 & 1 & 250$^{[1]}$ & 0.48 & $2.0\times 10^{-3}$ \\
%
\hline \hline 
\end{tabular}
%
\begin{flushleft}
{\footnotesize {
\textsuperscript{[1]}
$n_{\rm MTS}=30,100,200$ are also considered for benchmark calculations.
}} \\
{\footnotesize {
\textsuperscript{[2]}
The number of QM atoms.
}}    
\end{flushleft}
\end{table}


\subsection{Bulk Water}

We modelled bulk water by taking periodic cubic boxes containing 32 water molecules and 128 water molecules in two independent set of simulations. 
The cell edge lengths 
for 32 and 128 water systems were taken as $9.85$~{\AA} and $15.64$~{\AA} respectively, corresponding to the water density $\sim$1~g~cm$^{-3}$.

\subsection{\label{subsec:formamide}Formamide Solution}
A cubic periodic simulation cell with a side length of 10~{\AA} was chosen, which contained one formamide molecule, one hydroxide ion, and 29 water molecules.
%
%
%
%
%

\subsection{Benzoquinone Radical Anion in Methanol}
For the simulations of reduced $p$-benzoquinone radical (BQ$^{.-}$) in methanol solvent,
we took BQ$^{.-}$ radical and 57~MeOH molecules
in a periodic cubic box with 15.74~{\AA} 
side length, following the setup in
Ref. \citenum{BQ-ang}.
%
%
%
%
%
%
%
%
%

\subsection{Fe$^{3+}$ in Water}
We considered a configuration with a single Fe$^{3+}$ ion solvated in 64 water molecules inside a periodic cubic box of 12.414~{\AA} edge length, 
as in Ref. \citenum{sagar_Fe_CPC}.
%
%
BLYP functional was used in the {\bf GGA} run, while B3LYP functional was taken for {\bf VV} and {\bf RF-MTACE-$n_{\rm MTS}$} runs.
%
%

\subsection{TiO$_{2-x}$ Surface}
%
%
We modelled TiO$_{2-x}$ rutile(110) 
surface by taking a slab of $(3\times 2)$ Ti$_{48}$O$_{95}$ supercell with 4 trilayers.
%
A vacuum length of 10.000~\AA ~was introduced between the two slabs along $[110]$ direction,
and the supercell size was $21.974 \times 12.974 \times 8.862$~{\AA}$^3$ in our simulation. 
To model the defective surface with an oxygen vacancy, a 2-fold
coordinated oxygen atom was removed from the (110) surface. 
%
%
The optimized bulk structure of TiO$_2$ was adopted from Ref. \citenum{Marx_PRB_static_tio2}, with lattice parameters $a=b=4.649$~\AA, $c=2.966$~\AA, and $u=0.305$.
%
%
%
The calculations were performed at the $\Gamma$-point approximation.
%
%
%
%
%
%
%
%
%
%
%
%
%

\subsection{Enzyme:Drug Complex}
%
We carried out hybrid functional-based quantum mechanics/molecular mechanics (QM/MM) MD
simulation of acyl-enzyme covalent complex ({\bf EI}) of cephalothin and CBL.
The structure was built using 
AmpC with cephalothin structure from PDB ID 1KVM.
All the ionizable amino acids were set to their standard protonation state at pH $7$.
However, active site Lys$_{67}$ and Tyr$_{150}$ were taken in their neutral states based on our previous study.\cite{Ravi_2016_JPCB}
The AMBER parm99\cite{parm99} force field was used to describe the protein molecule.
Force-field parameters for cephalothin were obtained using the restrained electrostatic potential (RESP)\cite{RESP_2000} derived point charges and generalized AMBER force field (GAFF).\cite{gaff}
The system was solvated with 12808 TIP3P flexible water molecules in a periodic box of dimensions $81\times 88\times 76$~{\AA}$^3$, yielding a total of 44040 atoms.
Two Cl$^-$ atoms were added to neutralize the charge of the system.
Molecular mechanics (MM) calculations were carried out using the AMBER~18 package.\cite{amber18}
A total of $10000$ steps of energy minimization were performed, 
%
followed by equilibration at $300 \, \rm K$ and $1$ atm pressure for $\sim1$ ns till the density was converged.
Eventually, the system was equilibrated in an $NVT$ ensemble until a reasonable convergence in the protein backbone root mean square deviation (RMSD) was observed.
The equilibrated {\bf EI} structure thus obtained was then used to perform QM/MM MD simulations.

The performance benchmark of QM/MM canonical ensemble ($NVT$) {\bf RF-MTACE}-$n_{\rm MTS}$ runs were carried out using the CPMD/GROMOS interface\cite{Laio_2002_JCP, D-RESP, Gromos96_package,Gromos96_note} as implemented in the CPMD package.~\cite{cpmd}
The QM subsystem comprised the side chains of the active site residues, namely Ser64, Lys67, Lys315, and Tyr150, the hydrolytic water molecule, and the cephalothin drug molecule. 
The rest of the protein and solvent formed the MM subsystem and were treated using the GROMOS force field.~\cite{Gromos96_package, Gromos96_note}
Covalent bonds cleaved at the QM/MM boundary were capped with H atoms.
A total of 76 atoms with a net charge of 1 formed a part of the QM supercell with side length 24~{\AA}.
The QM part was treated using KS-DFT and PW.
%
The PBE and PBE0 XC functionals were used together with
norm-conserving Troullier-Martin type pseudopotentials,\cite{PRB_TM}
%
and a PW cutoff of 70~Ry was used.
The QM-MM interactions were treated using the scheme proposed by Laio {\it et al.}~\cite{Laio_2002_JCP,D-RESP}
Any interactions within 15~{\AA} of the QM density were treated explicitly, whereas those beyond this range were treated using a multipole expansion of charge density up to the quadrupole term.
QM/MM BOMD simulations were carried out at 300K.
As a massive thermostatting is required for SIN(R), we replaced  bond-constraints by bond-restraints.
%
%
\subsection{Formamide Hydrolysis  in Basic Solution}
\begin{figure}
    \centering
    \includegraphics[scale=0.4]{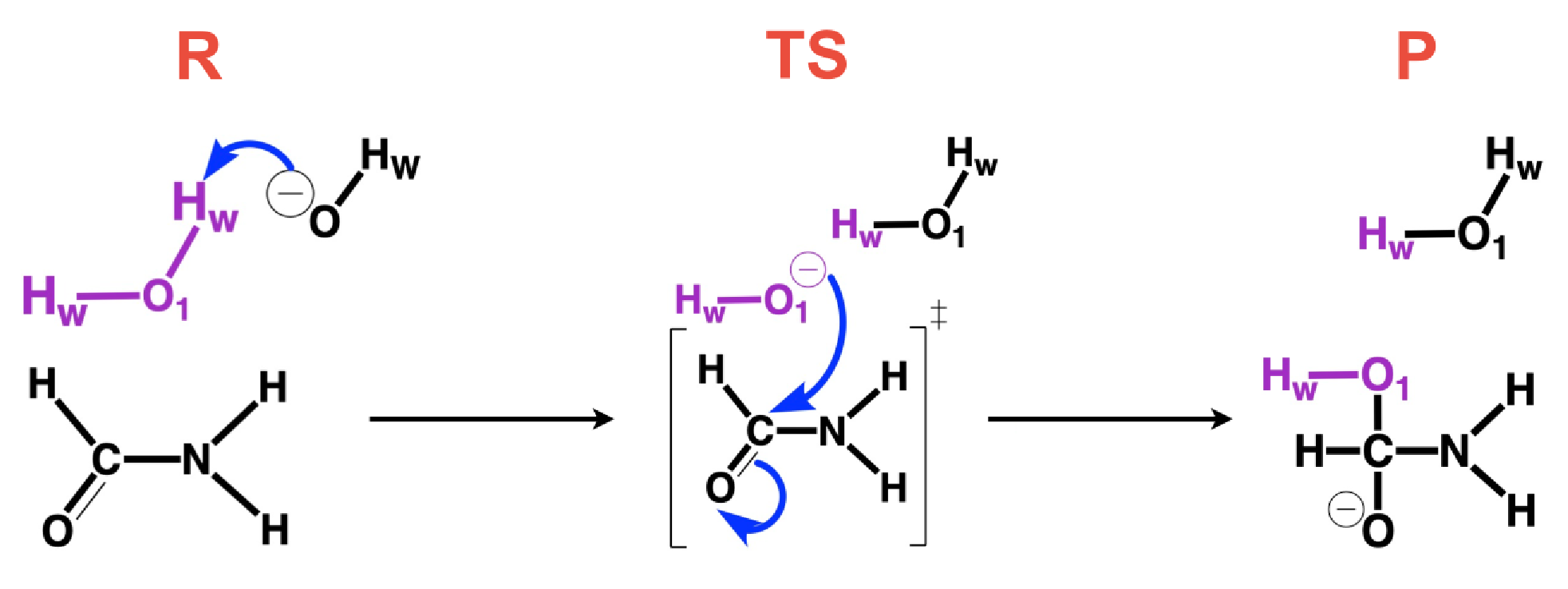}
    \caption{Formation of the tetrahedral intermediate ({\bf P}) from the reactant ({\bf R}) through the transition state ({\bf {TS}}) during formamide hydrolysis in basic solution.}
    \label{fig:reaction}
\end{figure}
%
%
We modelled the hydrolysis of formamide in basic solution, as depicted in \fref{fig:reaction}.
The system setup is identical to Section~\ref{subsec:formamide}.
To compute the free energy surface for the reaction, 
we used the well-sliced metadynamics (WS-MTD) technique.\cite{JCC_shalini} 
%
We chose two collective variables (CVs): (a)
$d$[C--O$_1$], distance between the carbon atom (C) of the formamide and the oxygen atom (O$_1$) of the attacking water molecule; (b)
coordination number ($CN$) of the oxygen atom (O$_1$) of the attacking water with all the hydrogen atoms (H$_{\textrm w}$) of the solvent molecules.
The $CN$ CV is defined as,
\begin{equation}
\label{CN}
CN[{\mathrm {O_1:H_w}}]=\sum_{i=1}^{N_{\mathrm {H_w}}} \frac{ 1}{1+({d_{1i}}/{d_0})^{6}} \enspace ,
\end{equation}
where $d_0=1.30$~{\AA}.
Here, $N_{\mathrm {H_w}}$ is the total number of H$_{\textrm w}$ atoms and $d_{1i}$ is the distance between the O$_1$ atom and the $i$-th H$_{\textrm w}$ atom.
%
%
The $d$[C--O$_1$] CV was sampled using the umbrella sampling\cite{US_method} like 
bias potential,
while
a well-tempered metadynamics (WT-MTD) bias potential\cite{WT-MTD}
was applied along the $CN$.
%

%
%
%
%
%
A total of 29 umbrella biasing windows were placed along $d$[C--O$_1$] in the range of 1.51 to 3.70~{\AA}.
%
%
%
%
For each umbrella window, we 
performed 50~ps of {\bf RF-MTACE}-$n_{\rm MTS}$ simulations, where only the last 40~ps trajectories were  considered for the analysis. 
%
Thus, we performed {\bf RF-MTACE}-$n_{\rm MTS}$ to generate $29\times 50~\mathrm{ps} = 1.45$~ns long trajectory at the hybrid-DFT level.
%
%
%
%
%

\section{Results and Discussion}
\subsection{Performance of RF-MTACE Method}
\begin{table}[h]
\caption{\label{tab:L2-speedup} Outer time step ($\Delta t$), $L^2$ error of RDFs, length of the simulation, average computational time per MD step ($t_{\rm MD}$), and speed-up for various runs.  
%
%
}
\centering
\begin{tabular}{|c|c|c|c|c|c|c|c|}
\hline \hline
Methods & $\Delta t$ (fs)  & $L_{g_{\rm OO}}^2$ & $L_{g_{\rm OH}}^2$ & $L_{g_{\rm HH}}^2$ & Length (ps) & $t_{\rm MD}$ (s) & Speed-up\\
\hline
{\bf VV} & 0.48 & 0.00 & 0.00 & 0.00  & 20 & 112.8 & 1 \\
\hline
{\bf MTACE}-15 & 7.2 & 0.03 & 0.04 & 0.04  & 30 & 11.2 & 10 \\
\hline
{\bf RF-MTACE}-30 & 14.4 & 0.05 & 0.04 & 0.05 & 30 & 7.0 & 16 \\
\hline
{\bf RF-MTACE}-100 & 48.4 & 0.04 & 0.03 & 0.03  & 30 & 4.4 & 26 \\
\hline
{\bf RF-MTACE}-200 & 96.8 & 0.05 & 0.04 & 0.03  & 30 & 3.8 & 30 \\
\hline
{\bf RF-MTACE}-250 & 120.0 & 0.04 & 0.04 & 0.03  & 30 & 3.7 & 31 \\ 
%
\hline \hline 
\end{tabular}
\end{table}

\begin{figure}
\includegraphics[scale=0.38]{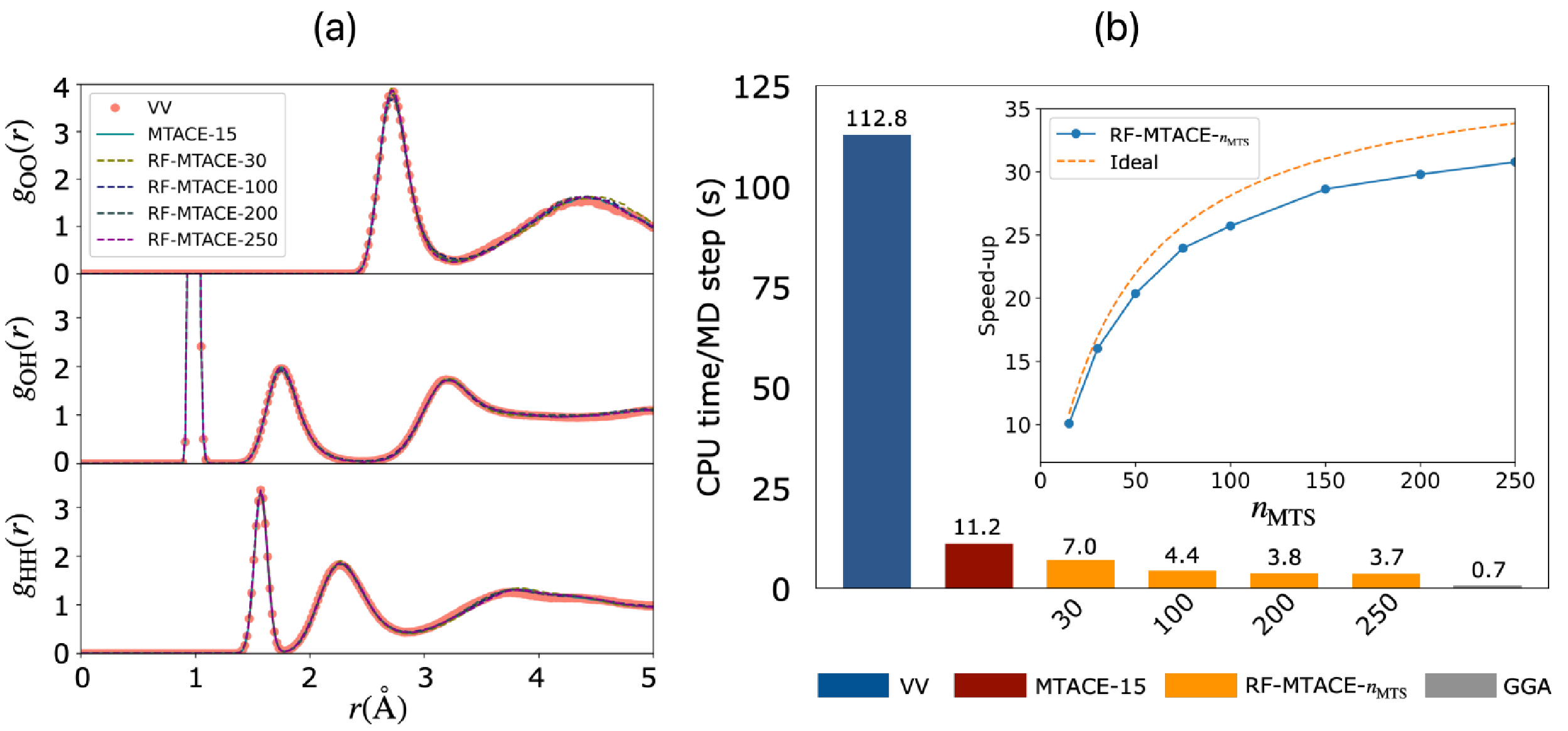}
\caption{\label{fig:rdf_scaling}
(a) O-O, O-H and H-H radial distribution functions (RDFs) for bulk 32 water periodic system during {\bf VV}, {\bf MTACE}-15 and {\bf {RF-MTACE-$n_{\rm MTS}$}} (with different time step factor $n_{\rm MTS}$) runs. 
%
%
(b) Average computational time per MD step with PBE0 and PBE functional for periodic 32 water model.
All the calculations were performed with 120 compute cores.
Ideal and actual speed-up for the {\bf RF-MTACE} runs (with respect to {\bf VV}) as a function of $n_{\rm MTS}$ are shown in the inset.
Ideal speed-up is defined as \eref{ideal-s}.
}
\end{figure}

We demonstrated the accuracy of the
RF-MTACE method by predicting the structural properties, in particular, the radial distribution functions (RDFs) of the bulk water system. 
These simulations were performed using the periodic 32-water system.
%
%
%
%
%
%
In Figure~\ref{fig:rdf_scaling}(a), the RDFs computed from the {\bf {RF-MTACE-$n_{\rm MTS}$}} runs with $n_{\rm MTS}$ values of 30, 100, 200, and 250 were compared with that of the {\bf VV} and {\bf MTACE}-15 runs.
The comparison of O-O, O-H and H-H RDFs  
clearly shows that the position and the height of the peaks match quite well with each other. 
We computed the $L^2$ error 
of the RDFs with reference to those obtained from {\bf VV} runs for a quantitative comparison, and the results are given in Table~\ref{tab:L2-speedup}.
The $L^2$ errors of RDFs from {\bf {RF-MTACE-$n_{\rm MTS}$}} runs 
for all the $n_{\rm MTS}$ values are remarkably small.
These results indicate that the choice of $n_{\rm MTS} > 15$ 
does not affect the structural properties of the system
when using the SIN(R) thermostat.
%
%
%

%
Although, the structural properties are reproduced well, we find that the dynamic properties are affected in the {\bf {RF-MTACE-$n_{\rm MTS}$}} runs.
It is known that the usage of SIN(R) thermostat will have an impact on the dynamic properties.\cite{SINR_main,SINR_polarised,SINR_MP_21}

%
%
We computed the speed-up of {\bf {RF-MTACE-$n_{\rm MTS}$}} runs compared to the {\bf VV} run using the following equation:
\begin{equation}
\label{ideal-s}
    {\mathrm {speed-up}}=\frac{t_{\rm VV}}{t_{\rm RF-MTACE-n_{\rm MTS}}} \enspace ,
\end{equation}
where $t_{\rm VV}$ is CPU time per MD step for the {\bf VV} run
and $t_{\rm RF-MTACE-n_{\rm MTS}}$ is CPU time per MD step for {\bf RF-MTACE}-$n_{\rm MTS}$ runs on identical number of compute cores.
%
%
Effective speed ups obtained in {\bf RF-MTACE}-30, {\bf RF-MTACE}-100, {\bf RF-MTACE}-200 and {\bf RF-MTACE}-250 runs are 
16, 26, 30 and 31, respectively, using identical 120~compute cores (Intel\textsuperscript{\textregistered} Skylake Xeon Platinum 8174 processors).  
The speed-up obtained in {\bf RF-MTACE}-$n_{\rm MTS}$ for different $n_{\rm MTS}$ values are given in \tref{tab:L2-speedup} and in \fref{fig:rdf_scaling}(b).
Notably, the previously reported {\bf MTACE}-15 method\cite{sagar_JCTC,sagar_JCC_scaling} could  offer only a speed up of 10 for the same test system.
The computational cost to perform one H-AIMD step with {\bf RF-MTACE}-250 is 3.7~s, which is now only 5 times slower than that of a {\bf GGA} run.

In \fref{fig:rdf_scaling}(b), we have also plotted the ideal speed up, where
$t_{\rm RF-MTACE-n_{\rm MTS}}$ in \eref{ideal-s} is calculated as
%
%
%
\begin{equation}
  t_{\rm RF-MTACE-n_{\rm MTS}} = \frac{t_{\rm exact}^{\rm force} + n_{\rm MTS} t_{\rm ACE}^{\rm force}}{n_{\rm MTS}}  \enspace .
\end{equation}
Here, $t_{\rm exact}^{\rm force}$ and $t_{\rm ACE}^{\rm force}$ are the average computational time for ${\textbf F}^{\textrm{exact}}$ and ${\textbf F}^{\textrm{ACE}}$ force calculations, respectively.
%
%
%
This equation considers the fact 
that $\mathbf F^{\rm ACE}$ 
is computed $n_{\rm MTS}$ times, whereas $\mathbf F^{\rm exact}$ is computed only once in every $n_{\rm MTS}$ MD steps.
%
%
%
Ideal speed-up and actual speed-up as a function of $n_{\rm MTS}$ are plotted in Figure~\ref{fig:rdf_scaling}(b).
We observe that the actual speed-up deviates from the ideal case, which is due to the increase in the number of SCF iterations when using large $n_{\rm MTS}$. 
%
%
Although, $n_{\rm MTS}=250$ could correctly reproduce the structural properties, the effect on the speed-up is not significant beyond $n_{\rm MTS}=200$. 
%
%
%
%

\subsection{Scaling Performance of RF-MTACE with Task Group Implementation}

\begin{figure}
\includegraphics[scale=0.5]{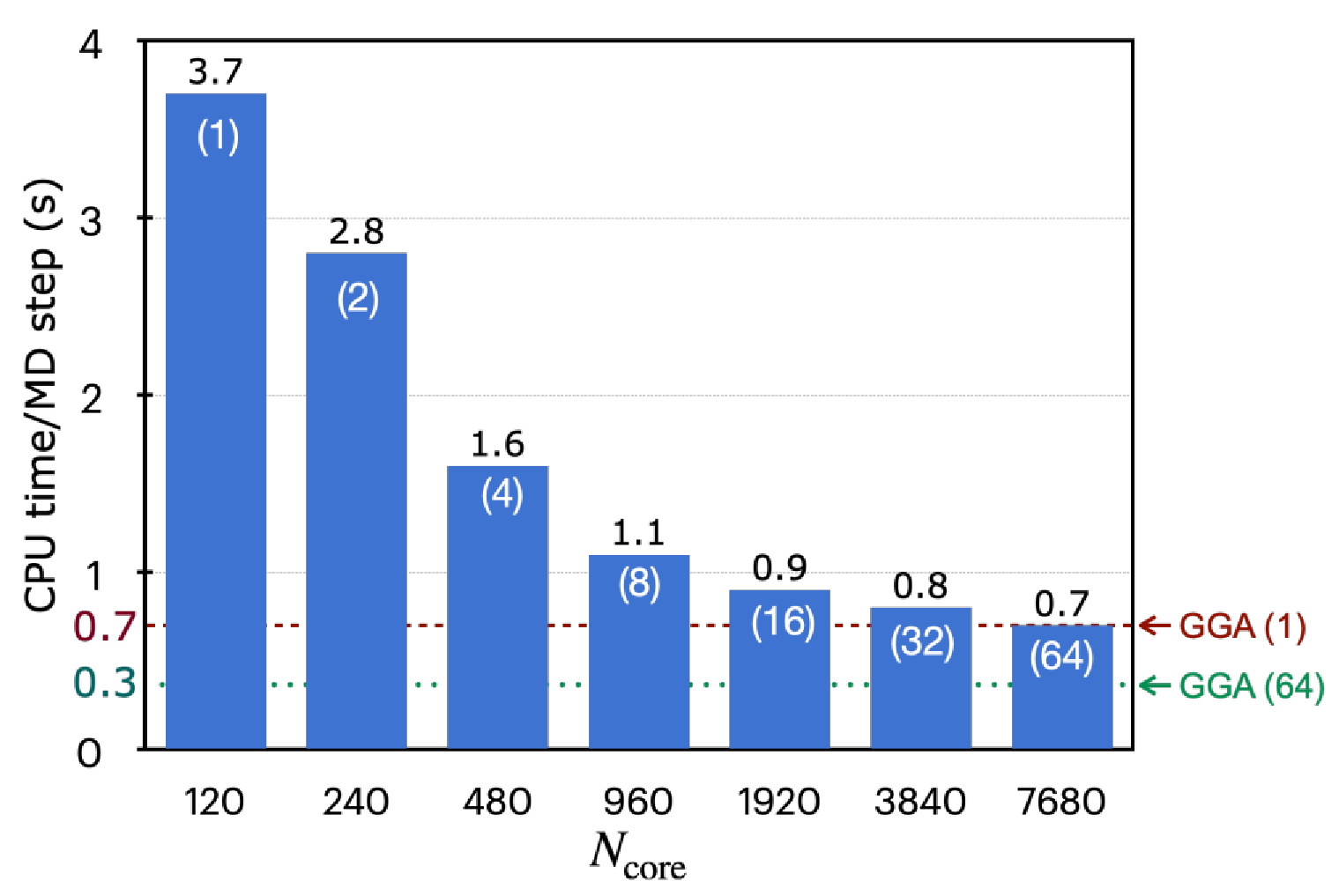}
\caption{\label{fig:cpg_result} Compute time per MD step for 32-water system in {\bf RF-MTACE}-250 for different $N_{\rm core}$ and $N_{\rm CPG}$. The $N_{\rm CPG}$ values are given within the parenthesis. Compute time per MD step in {\bf GGA} runs using 120~cores is indicated by the maroon dashed horizontal line, 
while the green dotted line indicates the compute time per MD step using 7680 compute cores.
}
\end{figure}

\begin{table}[h]
\caption{\label{cpg-tab}Average computational time per MD step for 32-water system using {\bf RF-MTACE}$-250$ and {\bf VV} for varying $N_{\rm core}$ and $N_{\rm CPG}$.
%
%
Speed-up is calculated using \eref{ideal-s}
with respect to {\bf VV}
using the same $N_{\rm core}$ and $N_{\rm CPG}$.}
\centering
\begin{tabular}{|c c c c c|}
\hline \hline
$N_{\rm core}$ & $N_{\rm CPG}$ & $t_{\rm VV}$(s) & $t_{\rm RF-MTACE-250}$(s) & Speed-up\\
\hline
120 & 1 & 110.2 & 3.7 & 30 \\
%
240 & 2 & 77.9 & 2.8 & 28 \\
480 & 4 & 37.8 & 1.6 & 24 \\
960 & 8 & 27.6 & 1.1 & 25 \\
1920 & 16 & 14.2 & 0.9 & 16 \\
3840 & 32 & 8.0 & 0.8 & 10 \\
7680 & 64 & 3.6 & 0.7 & 5 \\
%
\hline \hline 
\end{tabular}
\end{table}

So far, we have discussed the performance of the proposed method employing only 120 compute cores, which is the optimal number of cores that can be used for the 32-water system, i.e., $N_{\rm core}=N_{x}$.
Now, we employ the {\cpgroup} implementation to check the performance of {\bf RF-MTACE}-250 for
$N_{\rm core}>N_{x}$. 
The results are presented in \fref{fig:cpg_result} and \tref{cpg-tab}.
%
%
%
We find that compute time per MD step become nearly identical to that of the GGA with increasing number of compute cores to 7680 and using $N_{\rm CPG}=64$.
It has to be noted that {\bf VV} runs also scale well with the {\cpgroup} implementation. 
With 7680 compute cores,  the {\bf VV} run is only 
5 times slower than {\bf RF-MTACE}$-250$.
Due to this reason, the speed-up is decreasing with increasing $N_{\rm core}$, although the compute time decreases significantly.
%
%
While it is possible to obtain additional speed-up for {\bf GGA} with $N_{\rm core}>N_{x}$ using {\cpgroup}, the improvement is not significant.
In this case, we find that {\bf RF-MATCE}$-250$ run is only $\sim 1.5$ times slower than {\bf GGA} for $N_{\rm core}=7680$ and $N_{\rm CPG}=64$.
%
This is a significant performance enhancement. 
%

%
%

\subsection{Performance of RF-MTACE for Various Systems}
\begin{table}[h]
\caption{\label{systems} Average computational time per MD step for different systems using {\bf GGA}, {\bf VV} and {\bf RF-MTACE}$-n_{\rm MTS}$.
%
%
Speed-up is calculated using \eref{ideal-s}.
See \tref{tab:systems} for $N_{\rm at}$, $n_{\rm core}$, $N_{\rm CPG}$ and $n_{\rm MTS}$.
%
}
\centering
\begin{tabular}{|l|c|c|c|c|}
\hline \hline
System & $t_{\rm GGA}$(s) & $t_{\rm VV}$(s) & $t_{\rm RF-MTACE}$(s) & Speed-up\\
\hline
 128 water & 9.3 & 5499.8 & 168.6 & 33\\
\hline
Formamide solution & 0.9 & 130.6 & 5.8 & 23 \\
\hline
BQ$^{.-}$ in MeOH & 15.1 & 15855.2 & 314.8 & 50 \\
\hline
Fe$^{3+}$ in water & 9.8 & 1835.1 & 58.4 & 31 \\
\hline
TiO$_{2-x}$ surface & 64.2 & 8022.2 & 206.8 & 39 \\
\hline
Enzyme:Drug & 16.4 & 945.7 & 71.6$^{[1]}$ & 13 \\ 
\hline \hline 
\end{tabular}
%
\begin{flushleft}
{\footnotesize {
\textsuperscript{[1]}
Best performance for QM/MM was observed for $n_{\rm MTS}=250$.
}}    
\end{flushleft}
\end{table}

Having demonstrated the enhanced performance of RF-MTACE with the 32 water model system,
we turn our attention to a diverse set of  systems with particular significance.
%
%
These systems have the potential to benefit considerably from hybrid functional-based simulations in terms of improved predictive capabilities compared to {\bf GGA}. 
%
%
%
For each of these systems, we have reported the average computing time per MD step in {\bf GGA}, {\bf RF-MTACE}$-n_{\rm MTS}$, and {\bf VV} runs, as mentioned in \tref{systems}. 
%
%

%
It is known  that the structure and dynamics of liquid water, as predicted by the hybrid functionals, are in better agreement with the experimental results as compared to the GGA functional based results.\cite{JPCB_AIMD_HFX,JCTC_AIMD_HFX,JCP_AIMD_HFX,Mol_Phy_Car_MLWF,Mol_Phy_Car_MLWF_1,JPCB_water_hfx} 
%
%
%
%
From Table~\ref{systems}, we found that we could achieve a speed-up of $\sim$33 for 128-water system (384 atoms).
%
%
%

%
%
%
%
On the other hand, for the formamide solution system containing 95~atoms, the {\bf RF-MTACE}-200 run was found to be 23 times faster than the conventional {\bf VV} simulation (see Table~\ref{systems}). 
{It was observed that the number of SCF iterations in ${\textbf F}^{\textrm{exact}}$ force computations for {\bf RF-MTACE}-200 is higher than that of {\bf VV}. This explain the lower speed-up compared to 32 water system. 
Moreover, the smaller system size also contributes to the observed lower speed-up.}
%
%

%
Next, we focus on an open shell system where unrestricted DFT calculations are required.
Quinone derivatives play crucial role in various biologically important processes, such as photosynthesis.\cite{quinone_2016,quinone_PNAS,quinone_JACS,BQ-expt1,BQ-expt2} 
In this respect, redox properties of $p-$benzoquinone in different solvents have been studied.\cite{BQ-ang} 
%
%
%
%
%
%
%
%
For the $p-$benzoquinone in methanol system, we achieved a speed-up of 50 employing the RF-MTACE method (Table~\ref{systems}).
%
%
%
%

%
Another redox system of considerable interest is Fe$^{3+}$ ion in water.\cite{Fe3_IC,Fe3_JPCB,Cu_solv_Sc,sagar_Fe_CPC}
%
%
%
%
Here, we could achieve a speed-up of 31 with the help of the RF-MTACE method; see Table~\ref{systems}.
%
%

%
%
%
One of the most prevalent point defects on the TiO$_2$ rutile surface is oxygen vacancies, which significantly influence properties and reactivity of the material.\cite{exp_tio2_1_nat,exp_tio2_2_sc}
GGA functionals fail to describe the localized electronic state accurately and tend to underestimate the band gap,\cite{JCP_B3LYP,JCP_PBE0,JCP_PBE0_model,JCP_HSE,PCCP_HFX,PCCP_HSE,Galli_RSB_JPCL,Chem_Rev_Cohen,jpcl_2011_hfx,jpcl_2019_hfx,jpcl_2020_hfx} and it is crucial to use hybrid functionals or DFT+{\it U}.\cite{Marx_PRB_static_tio2,Marx_PRL_dyn_tio2,Selloni_meoh_ads_tio2,wat_ads_tio2_selloni,tio2_hubbard_diff_surf,tio2_band_gap_hubbard}
%
%
%
%
%
%
We found that {\bf RF-MTACE}-200 runs, using $N_{\rm core}=960$ and $N_{\rm CPG}=4$, gives a speed-up of 39 compared to the {\bf VV} run (Table~\ref{systems}).
%
%
%

\begin{figure}[h]
\includegraphics[scale=1.5]{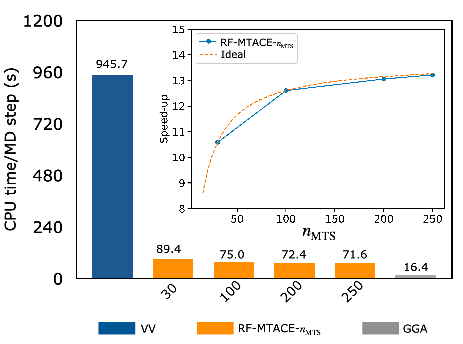}
\caption{\label{fig:QMMM_speed}
Average computational time per MD step with PBE0 and PBE functional for the QM/MM system.
All the calculations were performed with 240 compute cores.
Ideal and actual speed-up for the {\bf RF-MTACE} runs (with respect to {\bf VV}) as a function of $n_{\rm MTS}$ are shown in the inset.
Ideal speed-up is defined as \eref{ideal-s}.
}
\end{figure}
%
%
To assess the applicability of {RF-MTACE} method in QM/MM calculations, especially in the modeling of enzymatic reactions, we benchmarked the performance of {\bf RF-MTACE}-$n_{\rm MTS}$ runs for the acyl-enzyme complex of
CBL and cephalothin drug molecule.
This system is of great importance in the understanding of antibiotic resistance caused by the $\beta-$lactamase enzyme.
%
%
%
%
%
%
Our group has been actively working on understanding antibiotic resistance by different classes of $\beta-$lactamases.\cite{Ravi_2012_JPCB,Ravi_2013_JACS,Ravi_2016_JPCB,Chandan:pccp:2017,Chandan:ChemEurJ:2020,Shalini_AZT}
%
%
Despite their known limitation in underestimating the proton transfer barriers, GGA functionals are the common choice for studying these reactions at the QM/MM level.\cite{Adamo:2012}
%
%
Recently, we used the s-MTACE approach to compute the proton transfer barrier within the active site residues of CBL at the hybrid and GGA levels of functionals.\cite{sagar_JCTC}
In this study, it was found that the most stable protonation state is differently predicted by hybrid functional compared to GGA.
%
%
%
%
Here, we compared the performance of {\bf RF-MTACE}-$n_{\rm MTS}$ runs for the same system, as shown in \fref{fig:QMMM_speed}.
%
%
%
%
%
Our benchmark results with {\bf RF-MTACE}-200 and {\bf RF-MTACE}-250 runs suggest that the proposed method could speed up the calculations by a factor of 13 as compared to {\bf VV} (Table~\ref{systems}), which agrees well with the ideal speed-up (\fref{fig:QMMM_speed}).
The lower speed-up observed in this case can be attributed to the relatively smaller size of the QM system, which consists of only 76 atoms.
Additionally, the poor scaling of the MM and QM-MM electrostatics also contributes to the lower speed-up. 
%
%
%
%
%
%
%


\section{Application: Formamide Hydrolysis in Basic Solution}

\begin{figure}
\includegraphics[scale=0.4]{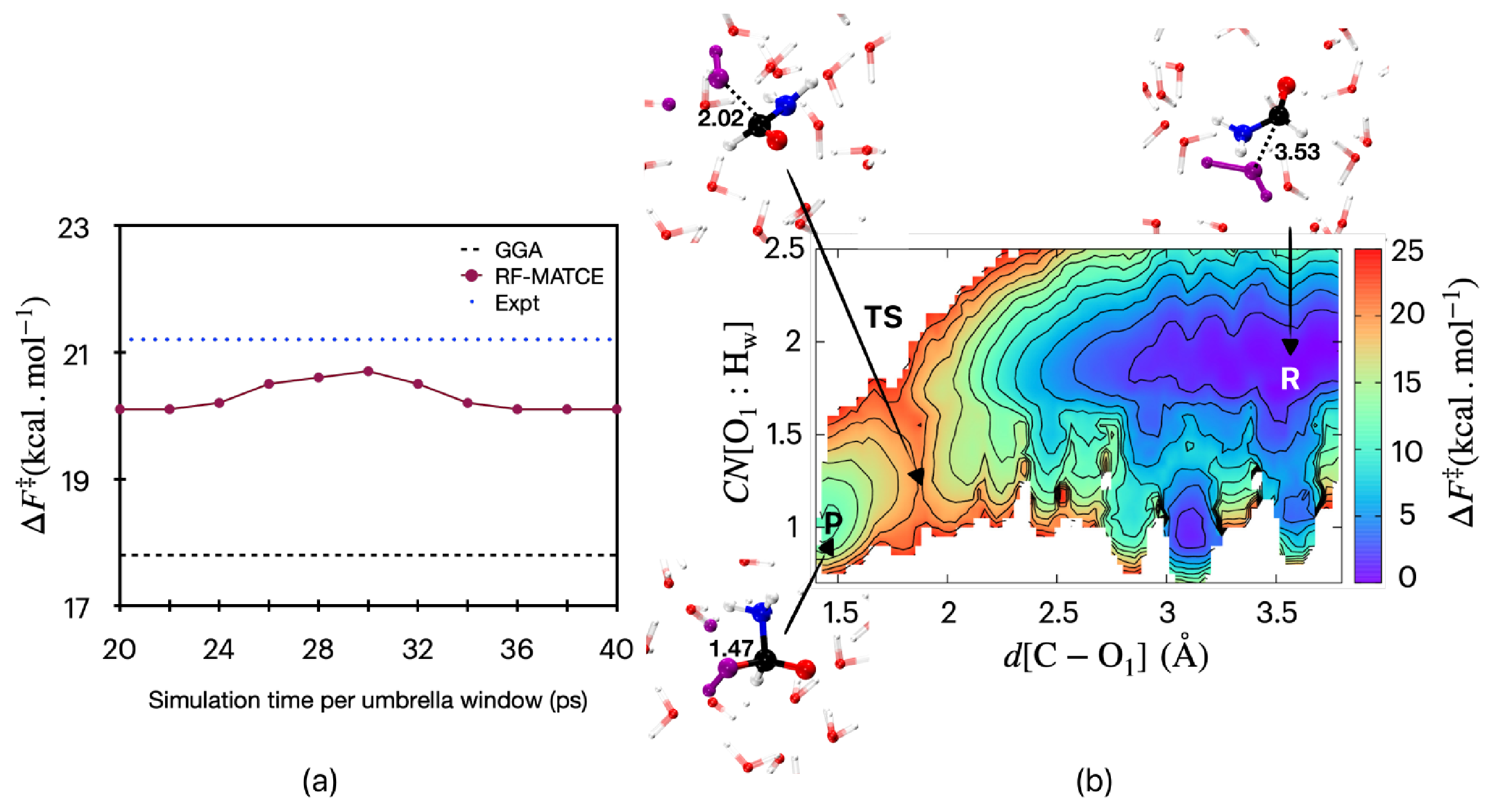}
\caption{\label{fig:free-energy}
(a) Convergence of free energy barrier of the formamide hydrolysis reaction in aqueous alkaline medium. 
The estimates for the free energy barrier from RF-MTACE simulation is compared with that of the {\bf GGA} calculations\cite{sagar_JCC} (black dashed line).
Experimental result is shown by the blue dotted line.\cite{Formamide_Exp}
%
%
(b) The reconstructed free energy surface from the {\bf RF-MTACE}-200 simulation. 
The contour lines are separated by 2 kcal/mol energy.
The snapshots of the tetrahedral intermediate ({\bf P}), the reactant ({\bf R}) and the transition state ({\bf TS}) are shown.
The attacking water is shown in violet.
Atom color codes: C (black), O (red), N (blue),  H (white), and attacking water molecule (violet).
The indicated distances are in {\AA}.
%
}
\end{figure}

Finally, as an application, we used the RF-MTACE method for computing the free energy surface of the formamide hydrolysis reaction in alkaline solution.  
This reaction is well studied experimentally and theoretically.\cite{Formamide_Exp,Angew_Klein,CPL_Klein}
In basic aqueous solution, the reaction proceeds via the formation of a tetrahedral intermediate.
%
%
To obtain the free-energy surface of the reaction, we employed WS-MTD technique.
%
%
From a total $29 \times 40$~ps of trajectories, we constructed the free energy surface using the reweighting procedure explained in Ref~\citenum{JCC_shalini}; see~\fref{fig:free-energy}(b). 
The free energy landscape agrees well with our previous studies.\cite{JCP_sagar,sagar_JCC}
To ensure the statistical convergence of the free energy estimates, we monitored free energy barrier as a function of simulation time per umbrella window, as shown in \fref{fig:free-energy}(a).
%
%
The final results were compared with those obtained using {\bf GGA}.\cite{JCP_sagar,sagar_JCC}
%
%
We observed that the converged free-energy barrier obtained from {\bf RF-MTACE}-200 simulation is 20.1~kcal~mol$^{-1}$, while using GGA, the barrier is 17.8~kcal~mol$^{-1}$.\cite{JCP_sagar} 
Notably, the barrier computed from {\bf RF-MTACE}-200 is also closer to the experimental result.\cite{Formamide_Exp}
%
%
Furthermore, the agreement in the computed barrier in RF-MTACE with previous studies suggests that the use of high time step factor has no effect on the free energy calculations.
%
%
%

%

%

\section{\label{sec:concl}Conclusions}

We have proposed a new variant of the MTACE method, named as RF-MTACE, which employs stochastic resonance-free thermostat together with multiple time stepping using ACE operator to speed-up hybrid density functional based AIMD simulations. 
%
%
%
The benchmark results show that the RF-MTACE method minimises computational expense substantially without compromising on the accuracy of static properties of the systems under consideration.
Usage of task group based parallelization 
allowed us to improve the performance of our method on large number of compute cores.
%
We achieved a speed-up to the extent that H-AIMD calculations are only marginally slower than the GGA level AIMD simulations.
%
%
%
%
%
%
Through 1.45~ns of H-AIMD simulations, we demonstrated that our technique could accurately predict the free-energy barrier of the hydrolysis of formamide in an alkaline solution.
%
We hope that our method will open the possibility of using hybrid density functional based {\it ab initio} calculations for studying the properties of various condensed matter systems and modeling complex physiochemical processes.

\newcounter{Aequ}
\newenvironment{AEquation}
  {\stepcounter{Aequ}%
    \addtocounter{equation}{-1}%
    \renewcommand\theequation{A\arabic{Aequ}}\equation}
  {\endequation}

\newcounter{Bequ}
\newenvironment{BEquation}
  {\stepcounter{Bequ}%
    \addtocounter{equation}{-1}%
    \renewcommand\theequation{B\arabic{Bequ}}\equation}
  {\endequation}
  
\newcounter{Cequ}
\newenvironment{CEquation}
  {\stepcounter{Cequ}%
    \addtocounter{equation}{-1}%
    \renewcommand\theequation{C\arabic{Cequ}}\equation}
  {\endequation}
\section*{Appendix}
\label{appendix}

The equations obtained after the application of \eref{trotter-sinr} on phase space variables are given here. \\
%
%
{\bf A : Update $V_I,v_{1,I,j},v_{2,I,j}$}

If $n_{\rm sy}$ is the Suzuki-Yoshida order and $n_{\rm res}$ is order of the RESPA decomposition of $e^{i{\mathcal L_{\rm N}}\delta t /2}$ operator in Eqn. 24, then,\cite{SINR_main}
\begin{AEquation}\label{equ:1}
    e^{i{\mathcal L_{\rm N}}\delta t /2}=\prod_{i=1}^{n_{\rm res}}\prod_{k=1}^{n_{\rm sy}}e^{i{\mathcal L_{\rm N}}\omega_k\delta t /2n_{\rm res}} \enspace .
\end{AEquation}
We denote, ${\Delta}=\omega_k\delta t/n_{\rm res}$. 
Application of $e^{i{\mathcal L_{\rm N}}\omega_k\delta t /2n_{\rm res}}$ on phase space variables are given below for a given $I=1,\cdots,3N_{\rm at}$ and $ k=1,\cdots,n_{\rm sy}$: 
%
\begin{AEquation}\label{equ:2}
\begin{split}
G(v_{1,I,j}) & = Q_1 v_{1,I,j}^2-k_B T \\
v_{2,I,j} & = v_{2,I,j}+\frac{{\Delta}}{4}\frac{G(v_{1,I,j})}{Q_2} ,\enspace \enspace j=1,\cdots,L \enspace .
\end{split}
\end{AEquation}
Now,
\begin{AEquation}\label{equ:3}
\begin{split}
V_I & =V_I H_I ,\\
v_{1,I,j} & =v_{1,I,j} H_I \exp({-v_{2,I,j}{\Delta}/2}) ,\enspace \enspace j=1,\cdots,L \enspace ,
\end{split}
\end{AEquation}
where,
\begin{AEquation}\label{equ:4}
    H_I =\sqrt{\frac{\Lambda_I}{M_I V_I^2 + \frac{L}{L+1}\sum_{j=1}^L Q_1 v_{1,I,j}^2 \exp({-v_{2,I,j}{\Delta}})}} \enspace .
\end{AEquation}
Again,
\begin{AEquation}\label{equ:4}
\begin{split}
G(v_{1,I,j}) & = Q_1 v_{1,I,j}^2-k_B T \enspace ,\\
v_{2,I,j} & = v_{2,I,j}+\frac{{\Delta}}{4}\frac{G(v_{1,I,j})}{Q_2},\enspace \enspace j=1,\cdots,L
\end{split}
\end{AEquation}
{\bf B : Update $V_I,v_{1,I,j}$}

\begin{BEquation}\label{equ:1}
\begin{split}
V_I & =\frac{V_I+sF_I/M_I}{\dot{s}} \enspace ,\\
v_{1,I,j} & = \frac{v_{1,I,j}}{\dot{s}} , \enspace \enspace j=1,\cdots,L 
\end{split}
\end{BEquation}
where,
\begin{BEquation}\label{equ:2}
\begin{split}
s & = \frac{1}{\sqrt{b}}\sinh (\sqrt{b}\delta t/2)+\frac{a}{b}(\cosh (\sqrt{b}\delta t/2)-1) \\
\dot{s} & = \cosh (\sqrt{b}\delta t/2) + \frac{a}{\sqrt{b}}\sinh (\sqrt{b}\delta t/2)
\end{split}
\end{BEquation}
with
\begin{BEquation}\label{equ:3}
    a=\frac{F_IV_I}{\Lambda_I} \enspace, 
     \enspace \enspace
    b=\frac{F_I^2}{M_I\Lambda_I} \enspace .
\end{BEquation}
{\bf C : Update $X_I,v_{2,I,j}$}


\begin{CEquation}\label{equ:1}
\begin{split}
X_I & = X_I+V_I \delta t ,\\
v_{2,I,j} & = v_{2,I,j}e^{-\gamma \delta t}+\sigma R(\delta t)\sqrt{\frac{1-e^{-2\gamma \delta t}}{2\gamma}}  ,\enspace \enspace j=1,\cdots,L ,
\end{split}
\end{CEquation}
where, $R(\delta t)$ is a Gaussian random number.


\begin{acknowledgement}
%
Financial support from National Supercomputing Mission (Subgroup Materials and Computational Chemistry), from Science and Engineering Research Board (India) under the MATRICS (Ref. No. MTR/2019/000359), and from the German Research Foundation (DFG) through SFB 953 (project number 182849149) are gratefully acknowledged.
R.K. and V.T. thank the Council of Scientific \& Industrial Research (CSIR), India, and IITK for their PhD fellowships.
Computational resources were provided by 
SuperMUC-NG (project pn98fa) at Leibniz Supercomputing Centre (LRZ), Param Sanganak (IIT Kanpur), and Param Shivay (IIT BHU) under the National Supercomputing Mission, India.
\end{acknowledgement}



%


%


\bibliography{achemso-demo}

\providecommand{\latin}[1]{#1}
\makeatletter
\providecommand{\doi}
  {\begingroup\let\do\@makeother\dospecials
  \catcode`\{=1 \catcode`\}=2 \doi@aux}
\providecommand{\doi@aux}[1]{\endgroup\texttt{#1}}
\makeatother
\providecommand*\mcitethebibliography{\thebibliography}
\csname @ifundefined\endcsname{endmcitethebibliography}
  {\let\endmcitethebibliography\endthebibliography}{}
\begin{mcitethebibliography}{132}
\providecommand*\natexlab[1]{#1}
\providecommand*\mciteSetBstSublistMode[1]{}
\providecommand*\mciteSetBstMaxWidthForm[2]{}
\providecommand*\mciteBstWouldAddEndPuncttrue
  {\def\EndOfBibitem{\unskip.}}
\providecommand*\mciteBstWouldAddEndPunctfalse
  {\let\EndOfBibitem\relax}
\providecommand*\mciteSetBstMidEndSepPunct[3]{}
\providecommand*\mciteSetBstSublistLabelBeginEnd[3]{}
\providecommand*\EndOfBibitem{}
\mciteSetBstSublistMode{f}
\mciteSetBstMaxWidthForm{subitem}{(\alph{mcitesubitemcount})}
\mciteSetBstSublistLabelBeginEnd
  {\mcitemaxwidthsubitemform\space}
  {\relax}
  {\relax}

\bibitem[Koch and Holthausen(2001)Koch, and Holthausen]{Chemist's_Guide}
Koch,~W.; Holthausen,~M.~C. \emph{A Chemist's Guide to Density Functional
  Theory}; WILEY-VCH: New York, 2001\relax
\mciteBstWouldAddEndPuncttrue
\mciteSetBstMidEndSepPunct{\mcitedefaultmidpunct}
{\mcitedefaultendpunct}{\mcitedefaultseppunct}\relax
\EndOfBibitem
\bibitem[Tuckerman(2002)]{Tuckerman_2002_aimd}
Tuckerman,~M.~E. Ab initio molecular dynamics: basic concepts, current trends
  and novel applications. \emph{J. Phys.: Condens. Matter} \textbf{2002},
  \emph{14}, R1297--R1355\relax
\mciteBstWouldAddEndPuncttrue
\mciteSetBstMidEndSepPunct{\mcitedefaultmidpunct}
{\mcitedefaultendpunct}{\mcitedefaultseppunct}\relax
\EndOfBibitem
\bibitem[Iftimie \latin{et~al.}(2005)Iftimie, Minary, and
  Tuckerman]{tuckerman_aimd}
Iftimie,~R.; Minary,~P.; Tuckerman,~M.~E. Ab initio molecular dynamics:
  Concepts, recent developments, and future trends. \emph{Proc. Natl. Acad.
  Sci. U.S.A.} \textbf{2005}, \emph{102}, 6654--6659\relax
\mciteBstWouldAddEndPuncttrue
\mciteSetBstMidEndSepPunct{\mcitedefaultmidpunct}
{\mcitedefaultendpunct}{\mcitedefaultseppunct}\relax
\EndOfBibitem
\bibitem[Marx and Hutter(2009)Marx, and Hutter]{marx-hutter-book}
Marx,~D.; Hutter,~J. \emph{Ab Initio Molecular Dynamics: Basic Theory and
  Advanced Methods}; Cambridge University Press: Cambridge, 2009\relax
\mciteBstWouldAddEndPuncttrue
\mciteSetBstMidEndSepPunct{\mcitedefaultmidpunct}
{\mcitedefaultendpunct}{\mcitedefaultseppunct}\relax
\EndOfBibitem
\bibitem[Cohen \latin{et~al.}(2008)Cohen, Mori-S{\'a}nchez, and
  Yang]{Science_DFT_limitations}
Cohen,~A.~J.; Mori-S{\'a}nchez,~P.; Yang,~W. Insights into Current Limitations
  of Density Functional Theory. \emph{Science} \textbf{2008}, \emph{321},
  792--794\relax
\mciteBstWouldAddEndPuncttrue
\mciteSetBstMidEndSepPunct{\mcitedefaultmidpunct}
{\mcitedefaultendpunct}{\mcitedefaultseppunct}\relax
\EndOfBibitem
\bibitem[Perdew and Zunger(1981)Perdew, and Zunger]{PRB_SIC}
Perdew,~J.~P.; Zunger,~A. Self-interaction correction to density-functional
  approximations for many-electron systems. \emph{Phys. Rev. B} \textbf{1981},
  \emph{23}, 5048--5079\relax
\mciteBstWouldAddEndPuncttrue
\mciteSetBstMidEndSepPunct{\mcitedefaultmidpunct}
{\mcitedefaultendpunct}{\mcitedefaultseppunct}\relax
\EndOfBibitem
\bibitem[Mori-S\'anchez \latin{et~al.}(2008)Mori-S\'anchez, Cohen, and
  Yang]{PRL_DFT_errors}
Mori-S\'anchez,~P.; Cohen,~A.~J.; Yang,~W. Localization and Delocalization
  Errors in Density Functional Theory and Implications for Band-Gap Prediction.
  \emph{Phys. Rev. Lett.} \textbf{2008}, \emph{100}, 146401\relax
\mciteBstWouldAddEndPuncttrue
\mciteSetBstMidEndSepPunct{\mcitedefaultmidpunct}
{\mcitedefaultendpunct}{\mcitedefaultseppunct}\relax
\EndOfBibitem
\bibitem[Becke(1988)]{PRA_GGA_Becke}
Becke,~A.~D. Density-functional exchange-energy approximation with correct
  asymptotic behavior. \emph{Phys. Rev. A} \textbf{1988}, \emph{38},
  3098--3100\relax
\mciteBstWouldAddEndPuncttrue
\mciteSetBstMidEndSepPunct{\mcitedefaultmidpunct}
{\mcitedefaultendpunct}{\mcitedefaultseppunct}\relax
\EndOfBibitem
\bibitem[Lee \latin{et~al.}(1988)Lee, Yang, and Parr]{PRB_GGA_LYP}
Lee,~C.; Yang,~W.; Parr,~R.~G. Development of the {C}olle-{S}alvetti
  correlation-energy formula into a functional of the electron density.
  \emph{Phys. Rev. B} \textbf{1988}, \emph{37}, 785--789\relax
\mciteBstWouldAddEndPuncttrue
\mciteSetBstMidEndSepPunct{\mcitedefaultmidpunct}
{\mcitedefaultendpunct}{\mcitedefaultseppunct}\relax
\EndOfBibitem
\bibitem[Perdew \latin{et~al.}(1996)Perdew, Burke, and Ernzerhof]{PRL_GGA_PBE}
Perdew,~J.~P.; Burke,~K.; Ernzerhof,~M. Generalized Gradient Approximation Made
  Simple. \emph{Phys. Rev. Lett.} \textbf{1996}, \emph{77}, 3865--3868\relax
\mciteBstWouldAddEndPuncttrue
\mciteSetBstMidEndSepPunct{\mcitedefaultmidpunct}
{\mcitedefaultendpunct}{\mcitedefaultseppunct}\relax
\EndOfBibitem
\bibitem[Bao \latin{et~al.}(2018)Bao, Gagliardi, and Truhlar]{JPCL_SIE}
Bao,~J.~L.; Gagliardi,~L.; Truhlar,~D.~G. Self-Interaction Error in Density
  Functional Theory: An Appraisal. \emph{J. Phys. Chem. Lett.} \textbf{2018},
  \emph{9}, 2353--2358\relax
\mciteBstWouldAddEndPuncttrue
\mciteSetBstMidEndSepPunct{\mcitedefaultmidpunct}
{\mcitedefaultendpunct}{\mcitedefaultseppunct}\relax
\EndOfBibitem
\bibitem[Martin(2004)]{Martin-book}
Martin,~R.~M. \emph{Electronic Structure: Basic Theory and Practical Methods};
  Cambridge University Press: Cambridge, 2004\relax
\mciteBstWouldAddEndPuncttrue
\mciteSetBstMidEndSepPunct{\mcitedefaultmidpunct}
{\mcitedefaultendpunct}{\mcitedefaultseppunct}\relax
\EndOfBibitem
\bibitem[Becke(1993)]{JCP_B3LYP}
Becke,~A.~D. Density‐functional thermochemistry. {III.} \uppercase{T}he role
  of exact exchange. \emph{J. Chem. Phys.} \textbf{1993}, \emph{98},
  5648--5652\relax
\mciteBstWouldAddEndPuncttrue
\mciteSetBstMidEndSepPunct{\mcitedefaultmidpunct}
{\mcitedefaultendpunct}{\mcitedefaultseppunct}\relax
\EndOfBibitem
\bibitem[Perdew \latin{et~al.}(1996)Perdew, Ernzerhof, and Burke]{JCP_PBE0}
Perdew,~J.~P.; Ernzerhof,~M.; Burke,~K. Rationale for mixing exact exchange
  with density functional approximations. \emph{J. Chem. Phys.} \textbf{1996},
  \emph{105}, 9982--9985\relax
\mciteBstWouldAddEndPuncttrue
\mciteSetBstMidEndSepPunct{\mcitedefaultmidpunct}
{\mcitedefaultendpunct}{\mcitedefaultseppunct}\relax
\EndOfBibitem
\bibitem[Heyd \latin{et~al.}(2003)Heyd, Scuseria, and Ernzerhof]{JCP_HSE}
Heyd,~J.; Scuseria,~G.~E.; Ernzerhof,~M. Hybrid functionals based on a screened
  {C}oulomb potential. \emph{J. Chem. Phys.} \textbf{2003}, \emph{118},
  8207--8215\relax
\mciteBstWouldAddEndPuncttrue
\mciteSetBstMidEndSepPunct{\mcitedefaultmidpunct}
{\mcitedefaultendpunct}{\mcitedefaultseppunct}\relax
\EndOfBibitem
\bibitem[Todorova \latin{et~al.}(2006)Todorova, Seitsonen, Hutter, Kuo, and
  Mundy]{JPCB_AIMD_HFX}
Todorova,~T.; Seitsonen,~A.~P.; Hutter,~J.; Kuo,~I.-F.~W.; Mundy,~C.~J.
  Molecular Dynamics Simulation of Liquid Water: Hybrid Density Functionals.
  \emph{J. Phys. Chem. B} \textbf{2006}, \emph{110}, 3685--3691\relax
\mciteBstWouldAddEndPuncttrue
\mciteSetBstMidEndSepPunct{\mcitedefaultmidpunct}
{\mcitedefaultendpunct}{\mcitedefaultseppunct}\relax
\EndOfBibitem
\bibitem[Zhang \latin{et~al.}(2011)Zhang, Donadio, Gygi, and
  Galli]{JCTC_AIMD_HFX}
Zhang,~C.; Donadio,~D.; Gygi,~F.; Galli,~G. First Principles Simulations of the
  Infrared Spectrum of Liquid Water Using Hybrid Density Functionals. \emph{J.
  Chem. Theory Comput.} \textbf{2011}, \emph{7}, 1443--1449\relax
\mciteBstWouldAddEndPuncttrue
\mciteSetBstMidEndSepPunct{\mcitedefaultmidpunct}
{\mcitedefaultendpunct}{\mcitedefaultseppunct}\relax
\EndOfBibitem
\bibitem[DiStasio~Jr. \latin{et~al.}(2014)DiStasio~Jr., Santra, Li, Wu, and
  Car]{JCP_AIMD_HFX}
DiStasio~Jr.,~R.~A.; Santra,~B.; Li,~Z.; Wu,~X.; Car,~R. The individual and
  collective effects of exact exchange and dispersion interactions on the ab
  initio structure of liquid water. \emph{J. Chem. Phys.} \textbf{2014},
  \emph{141}, 084502\relax
\mciteBstWouldAddEndPuncttrue
\mciteSetBstMidEndSepPunct{\mcitedefaultmidpunct}
{\mcitedefaultendpunct}{\mcitedefaultseppunct}\relax
\EndOfBibitem
\bibitem[Santra \latin{et~al.}(2015)Santra, DiStasio~Jr., Martelli, and
  Car]{Mol_Phy_Car_MLWF}
Santra,~B.; DiStasio~Jr.,~R.~A.; Martelli,~F.; Car,~R. Local structure analysis
  in ab initio liquid water. \emph{Mol. Phys.} \textbf{2015}, \emph{113},
  2829--2841\relax
\mciteBstWouldAddEndPuncttrue
\mciteSetBstMidEndSepPunct{\mcitedefaultmidpunct}
{\mcitedefaultendpunct}{\mcitedefaultseppunct}\relax
\EndOfBibitem
\bibitem[Bankura \latin{et~al.}(2015)Bankura, Santra, DiStasio~Jr., Swartz,
  Klein, and Wu]{Mol_Phy_Car_MLWF_1}
Bankura,~A.; Santra,~B.; DiStasio~Jr.,~R.~A.; Swartz,~C.~W.; Klein,~M.~L.;
  Wu,~X. A systematic study of chloride ion solvation in water using van der
  {W}aals inclusive hybrid density functional theory. \emph{Mol. Phys.}
  \textbf{2015}, \emph{113}, 2842--2854\relax
\mciteBstWouldAddEndPuncttrue
\mciteSetBstMidEndSepPunct{\mcitedefaultmidpunct}
{\mcitedefaultendpunct}{\mcitedefaultseppunct}\relax
\EndOfBibitem
\bibitem[Ambrosio \latin{et~al.}(2016)Ambrosio, Miceli, and
  Pasquarello]{JPCB_water_hfx}
Ambrosio,~F.; Miceli,~G.; Pasquarello,~A. Structural, Dynamical, and Electronic
  Properties of Liquid Water: A Hybrid Functional Study. \emph{J. Phys. Chem.
  B} \textbf{2016}, \emph{120}, 7456--7470\relax
\mciteBstWouldAddEndPuncttrue
\mciteSetBstMidEndSepPunct{\mcitedefaultmidpunct}
{\mcitedefaultendpunct}{\mcitedefaultseppunct}\relax
\EndOfBibitem
\bibitem[Janesko and Scuseria(2008)Janesko, and Scuseria]{Scuseria:JCP:2008}
Janesko,~B.~G.; Scuseria,~G.~E. Hartree-{F}ock orbitals significantly improve
  the reaction barrier heights predicted by semilocal density functionals.
  \emph{J.~Chem.~Phys.} \textbf{2008}, \emph{128}, 244112\relax
\mciteBstWouldAddEndPuncttrue
\mciteSetBstMidEndSepPunct{\mcitedefaultmidpunct}
{\mcitedefaultendpunct}{\mcitedefaultseppunct}\relax
\EndOfBibitem
\bibitem[Mangiatordi \latin{et~al.}(2012)Mangiatordi, Br\'emond, and
  Adamo]{Adamo:2012}
Mangiatordi,~G.~F.; Br\'emond,~E.; Adamo,~C. \uppercase{DFT} and Proton
  Transfer Reactions: A Benchmark Study on Structure and Kinetics.
  \emph{J.~Chem.~Theory Comput.} \textbf{2012}, \emph{8}, 3082--3088\relax
\mciteBstWouldAddEndPuncttrue
\mciteSetBstMidEndSepPunct{\mcitedefaultmidpunct}
{\mcitedefaultendpunct}{\mcitedefaultseppunct}\relax
\EndOfBibitem
\bibitem[Adamo and Barone(1999)Adamo, and Barone]{JCP_PBE0_model}
Adamo,~C.; Barone,~V. Toward reliable density functional methods without
  adjustable parameters: \uppercase{T}he \uppercase{PBE0} model. \emph{J. Chem.
  Phys.} \textbf{1999}, \emph{110}, 6158--6170\relax
\mciteBstWouldAddEndPuncttrue
\mciteSetBstMidEndSepPunct{\mcitedefaultmidpunct}
{\mcitedefaultendpunct}{\mcitedefaultseppunct}\relax
\EndOfBibitem
\bibitem[Cramer and Truhlar(2009)Cramer, and Truhlar]{PCCP_HFX}
Cramer,~C.~J.; Truhlar,~D.~G. Density functional theory for transition metals
  and transition metal chemistry. \emph{Phys. Chem. Chem. Phys.} \textbf{2009},
  \emph{11}, 10757--10816\relax
\mciteBstWouldAddEndPuncttrue
\mciteSetBstMidEndSepPunct{\mcitedefaultmidpunct}
{\mcitedefaultendpunct}{\mcitedefaultseppunct}\relax
\EndOfBibitem
\bibitem[Janesko \latin{et~al.}(2009)Janesko, Henderson, and
  Scuseria]{PCCP_HSE}
Janesko,~B.~G.; Henderson,~T.~M.; Scuseria,~G.~E. Screened hybrid density
  functionals for solid-state chemistry and physics. \emph{Phys. Chem. Chem.
  Phys.} \textbf{2009}, \emph{11}, 443--454\relax
\mciteBstWouldAddEndPuncttrue
\mciteSetBstMidEndSepPunct{\mcitedefaultmidpunct}
{\mcitedefaultendpunct}{\mcitedefaultseppunct}\relax
\EndOfBibitem
\bibitem[Wan \latin{et~al.}(2014)Wan, Spanu, Gygi, and Galli]{Galli_RSB_JPCL}
Wan,~Q.; Spanu,~L.; Gygi,~F.; Galli,~G. Electronic Structure of Aqueous
  Sulfuric Acid from First-Principles Simulations with Hybrid Functionals.
  \emph{J. Phys. Chem. Lett.} \textbf{2014}, \emph{5}, 2562--2567\relax
\mciteBstWouldAddEndPuncttrue
\mciteSetBstMidEndSepPunct{\mcitedefaultmidpunct}
{\mcitedefaultendpunct}{\mcitedefaultseppunct}\relax
\EndOfBibitem
\bibitem[Cohen \latin{et~al.}(2012)Cohen, Mori-S\'anchez, and
  Yang]{Chem_Rev_Cohen}
Cohen,~A.~J.; Mori-S\'anchez,~P.; Yang,~W. Challenges for Density Functional
  Theory. \emph{Chem. Rev.} \textbf{2012}, \emph{112}, 289--320\relax
\mciteBstWouldAddEndPuncttrue
\mciteSetBstMidEndSepPunct{\mcitedefaultmidpunct}
{\mcitedefaultendpunct}{\mcitedefaultseppunct}\relax
\EndOfBibitem
\bibitem[Xiao \latin{et~al.}(2011)Xiao, Tahir-Kheli, and
  Goddard~III]{jpcl_2011_hfx}
Xiao,~H.; Tahir-Kheli,~J.; Goddard~III,~W.~A. Accurate Band Gaps for
  Semiconductors from Density Functional Theory. \emph{J. Phys. Chem. Lett.}
  \textbf{2011}, \emph{2}, 212--217\relax
\mciteBstWouldAddEndPuncttrue
\mciteSetBstMidEndSepPunct{\mcitedefaultmidpunct}
{\mcitedefaultendpunct}{\mcitedefaultseppunct}\relax
\EndOfBibitem
\bibitem[Zhao and Kulik(2019)Zhao, and Kulik]{jpcl_2019_hfx}
Zhao,~Q.; Kulik,~H.~J. Stable Surfaces That Bind Too Tightly: Can
  Range-Separated Hybrids or {DFT+U} Improve Paradoxical Descriptions of
  Surface Chemistry? \emph{J. Phys. Chem. Lett.} \textbf{2019}, \emph{10},
  5090--5098\relax
\mciteBstWouldAddEndPuncttrue
\mciteSetBstMidEndSepPunct{\mcitedefaultmidpunct}
{\mcitedefaultendpunct}{\mcitedefaultseppunct}\relax
\EndOfBibitem
\bibitem[Gerrits \latin{et~al.}(2020)Gerrits, Smeets, Vuckovic, Powell,
  Doblhoff-Dier, and Kroes]{jpcl_2020_hfx}
Gerrits,~N.; Smeets,~E. W.~F.; Vuckovic,~S.; Powell,~A.~D.; Doblhoff-Dier,~K.;
  Kroes,~G.-J. Density Functional Theory for Molecule–Metal Surface
  Reactions: When Does the Generalized Gradient Approximation Get It Right, and
  What to Do If It Does Not. \emph{J. Phys. Chem. Lett.} \textbf{2020},
  \emph{11}, 10552--10560\relax
\mciteBstWouldAddEndPuncttrue
\mciteSetBstMidEndSepPunct{\mcitedefaultmidpunct}
{\mcitedefaultendpunct}{\mcitedefaultseppunct}\relax
\EndOfBibitem
\bibitem[Frisch \latin{et~al.}(2004)Frisch, Trucks, Schlegel, Scuseria, Robb,
  Cheeseman, Montgomery, Jr., A., Vreven, Kudin, Burant, Millam, Iyengar,
  Tomasi, Barone, Mennucci, Cossi, Scalmani, Rega, Petersson, Nakatsuji, Hada,
  Ehara, Toyota, Fukuda, Hasegawa, Ishida, Nakajima, Honda, Kitao, Nakai,
  Klene, Li, Knox, Hratchian, Cross, Bakken, Adamo, Jaramillo, Gomperts,
  Stratmann, Yazyev, Austin, Cammi, Pomelli, Ochterski, Ayala, Morokuma, Voth,
  Salvador, Dannenberg, Zakrzewski, Dapprich, Daniels, Strain, Farkas, Malick,
  Rabuck, Raghavachari, Foresman, Ortiz, Cui, Baboul, Clifford, Cioslowski,
  Stefanov, Liu, Liashenko, Piskorz, Komaromi, Martin, Fox, Keith, Al-Laham,
  Peng, Nanayakkara, Challacombe, Gill, Johnson, Chen, Wong, Gonzalez, and
  Pople]{Pople2003}
Frisch,~M.~J.; Trucks,~G.~W.; Schlegel,~H.~B.; Scuseria,~G.~E.; Robb,~M.~A.;
  Cheeseman,~J.~R.; Montgomery,; Jr.,; A.,~J.; Vreven,~T.; Kudin,~K.~N.;
  Burant,~J.~C.; Millam,~J.~M.; Iyengar,~S.~S.; Tomasi,~J.; Barone,~V.;
  Mennucci,~B.; Cossi,~M.; Scalmani,~G.; Rega,~N.; Petersson,~G.~A.;
  Nakatsuji,~H.; Hada,~M.; Ehara,~M.; Toyota,~K.; Fukuda,~R.; Hasegawa,~J.;
  Ishida,~M.; Nakajima,~T.; Honda,~Y.; Kitao,~O.; Nakai,~H.; Klene,~M.; Li,~X.;
  Knox,~J.~E.; Hratchian,~H.~P.; Cross,~J.~B.; Bakken,~V.; Adamo,~C.;
  Jaramillo,~J.; Gomperts,~R.; Stratmann,~R.~E.; Yazyev,~O.; Austin,~A.~J.;
  Cammi,~R.; Pomelli,~C.; Ochterski,~J.~W.; Ayala,~P.~Y.; Morokuma,~K.;
  Voth,~G.~A.; Salvador,~P.; Dannenberg,~J.~J.; Zakrzewski,~V.~G.;
  Dapprich,~S.; Daniels,~A.~D.; Strain,~M.~C.; Farkas,~O.; Malick,~D.~K.;
  Rabuck,~A.~D.; Raghavachari,~K.; Foresman,~J.~B.; Ortiz,~J.~V.; Cui,~Q.;
  Baboul,~A.~G.; Clifford,~S.; Cioslowski,~J.; Stefanov,~B.~B.; Liu,~G.;
  Liashenko,~A.; Piskorz,~P.; Komaromi,~I.; Martin,~R.~L.; Fox,~D.~J.;
  Keith,~T.; Al-Laham,~M.~A.; Peng,~C.~Y.; Nanayakkara,~A.; Challacombe,~M.;
  Gill,~P.~M.~W.; Johnson,~B.; Chen,~W.; Wong,~M.~W.; Gonzalez,~C.;
  Pople,~J.~A. {Gaussian} 03. Gaussian, Inc.: Wallingford, CT, 2004\relax
\mciteBstWouldAddEndPuncttrue
\mciteSetBstMidEndSepPunct{\mcitedefaultmidpunct}
{\mcitedefaultendpunct}{\mcitedefaultseppunct}\relax
\EndOfBibitem
\bibitem[Chawla and Voth(1998)Chawla, and Voth]{JCP_HFX_Voth}
Chawla,~S.; Voth,~G.~A. Exact exchange in ab initio molecular dynamics: An
  efficient plane-wave based algorithm. \emph{J. Chem. Phys.} \textbf{1998},
  \emph{108}, 4697--4700\relax
\mciteBstWouldAddEndPuncttrue
\mciteSetBstMidEndSepPunct{\mcitedefaultmidpunct}
{\mcitedefaultendpunct}{\mcitedefaultseppunct}\relax
\EndOfBibitem
\bibitem[Sharma \latin{et~al.}(2003)Sharma, Wu, and Car]{MLWF_Car}
Sharma,~M.; Wu,~Y.; Car,~R. Ab initio molecular dynamics with maximally
  localized Wannier functions. \emph{Int. J. Quantum Chem.} \textbf{2003},
  \emph{95}, 821--829\relax
\mciteBstWouldAddEndPuncttrue
\mciteSetBstMidEndSepPunct{\mcitedefaultmidpunct}
{\mcitedefaultendpunct}{\mcitedefaultseppunct}\relax
\EndOfBibitem
\bibitem[Wu \latin{et~al.}(2009)Wu, Selloni, and Car]{PRB_Car_Wannier}
Wu,~X.; Selloni,~A.; Car,~R. Order-{$N$} implementation of exact exchange in
  extended insulating systems. \emph{Phys. Rev. B} \textbf{2009}, \emph{79},
  085102\relax
\mciteBstWouldAddEndPuncttrue
\mciteSetBstMidEndSepPunct{\mcitedefaultmidpunct}
{\mcitedefaultendpunct}{\mcitedefaultseppunct}\relax
\EndOfBibitem
\bibitem[Chen \latin{et~al.}(2018)Chen, Zheng, Santra, Ko, DiStasio~Jr., Klein,
  Car, and Wu]{Nature_Car_MLWF}
Chen,~M.; Zheng,~L.; Santra,~B.; Ko,~H.-Y.; DiStasio~Jr.,~R.~A.; Klein,~M.~L.;
  Car,~R.; Wu,~X. Hydroxide diffuses slower than hydronium in water because its
  solvated structure inhibits correlated proton transfer. \emph{Nat. Chem.}
  \textbf{2018}, \emph{10}, 413--419\relax
\mciteBstWouldAddEndPuncttrue
\mciteSetBstMidEndSepPunct{\mcitedefaultmidpunct}
{\mcitedefaultendpunct}{\mcitedefaultseppunct}\relax
\EndOfBibitem
\bibitem[Gygi(2009)]{PRL_RSB}
Gygi,~F. Compact Representations of {K}ohn-{S}ham Invariant Subspaces.
  \emph{Phys. Rev. Lett.} \textbf{2009}, \emph{102}, 166406\relax
\mciteBstWouldAddEndPuncttrue
\mciteSetBstMidEndSepPunct{\mcitedefaultmidpunct}
{\mcitedefaultendpunct}{\mcitedefaultseppunct}\relax
\EndOfBibitem
\bibitem[Gygi and Duchemin(2013)Gygi, and Duchemin]{JCTC_RSB}
Gygi,~F.; Duchemin,~I. Efficient Computation of {H}artree–{F}ock Exchange
  Using Recursive Subspace Bisection. \emph{J. Chem. Theory Comput.}
  \textbf{2013}, \emph{9}, 582--587\relax
\mciteBstWouldAddEndPuncttrue
\mciteSetBstMidEndSepPunct{\mcitedefaultmidpunct}
{\mcitedefaultendpunct}{\mcitedefaultseppunct}\relax
\EndOfBibitem
\bibitem[Dawson and Gygi(2015)Dawson, and Gygi]{JCTC_RSB_1}
Dawson,~W.; Gygi,~F. Performance and Accuracy of Recursive Subspace Bisection
  for Hybrid {DFT} Calculations in Inhomogeneous Systems. \emph{J. Chem. Theory
  Comput.} \textbf{2015}, \emph{11}, 4655--4663\relax
\mciteBstWouldAddEndPuncttrue
\mciteSetBstMidEndSepPunct{\mcitedefaultmidpunct}
{\mcitedefaultendpunct}{\mcitedefaultseppunct}\relax
\EndOfBibitem
\bibitem[Gaiduk \latin{et~al.}(2014)Gaiduk, Zhang, Gygi, and
  Galli]{Galli_RSB_CPL}
Gaiduk,~A.~P.; Zhang,~C.; Gygi,~F.; Galli,~G. Structural and electronic
  properties of aqueous {NaCl} solutions from ab initio molecular dynamics
  simulations with hybrid density functionals. \emph{Chem. Phys. Lett.}
  \textbf{2014}, \emph{604}, 89 -- 96\relax
\mciteBstWouldAddEndPuncttrue
\mciteSetBstMidEndSepPunct{\mcitedefaultmidpunct}
{\mcitedefaultendpunct}{\mcitedefaultseppunct}\relax
\EndOfBibitem
\bibitem[Gaiduk \latin{et~al.}(2015)Gaiduk, Gygi, and Galli]{Galli_RSB_JPCL1}
Gaiduk,~A.~P.; Gygi,~F.; Galli,~G. Density and Compressibility of Liquid Water
  and Ice from First-Principles Simulations with Hybrid Functionals. \emph{J.
  Phys. Chem. Lett.} \textbf{2015}, \emph{6}, 2902--2908\relax
\mciteBstWouldAddEndPuncttrue
\mciteSetBstMidEndSepPunct{\mcitedefaultmidpunct}
{\mcitedefaultendpunct}{\mcitedefaultseppunct}\relax
\EndOfBibitem
\bibitem[Mandal \latin{et~al.}(2018)Mandal, Debnath, Meyer, and
  Nair]{JCP_sagar}
Mandal,~S.; Debnath,~J.; Meyer,~B.; Nair,~N.~N. Enhanced sampling and free
  energy calculations with hybrid functionals and plane waves for chemical
  reactions. \emph{J. Chem. Phys.} \textbf{2018}, \emph{149}, 144113\relax
\mciteBstWouldAddEndPuncttrue
\mciteSetBstMidEndSepPunct{\mcitedefaultmidpunct}
{\mcitedefaultendpunct}{\mcitedefaultseppunct}\relax
\EndOfBibitem
\bibitem[Ko \latin{et~al.}(2020)Ko, Jia, Santra, Wu, Car, and
  DiStasio~Jr.]{Car_hfx_2019}
Ko,~H.-Y.; Jia,~J.; Santra,~B.; Wu,~X.; Car,~R.; DiStasio~Jr.,~R.~A. Enabling
  Large-Scale Condensed-Phase Hybrid Density Functional Theory Based Ab Initio
  Molecular Dynamics. 1. Theory, Algorithm, and Performance.
  \emph{J.~Chem.~Theory Comput.} \textbf{2020}, \emph{16}, 3757--3785\relax
\mciteBstWouldAddEndPuncttrue
\mciteSetBstMidEndSepPunct{\mcitedefaultmidpunct}
{\mcitedefaultendpunct}{\mcitedefaultseppunct}\relax
\EndOfBibitem
\bibitem[Ko \latin{et~al.}(2021)Ko, Santra, and DiStasio]{enabling_part2}
Ko,~H.-Y.; Santra,~B.; DiStasio,~R. A.~J. Enabling Large-Scale Condensed-Phase
  Hybrid Density Functional Theory-Based Ab Initio Molecular Dynamics II:
  Extensions to the Isobaric–Isoenthalpic and Isobaric–Isothermal
  Ensembles. \emph{J. Chem. Theory Comput.} \textbf{2021}, \emph{17},
  7789--7813\relax
\mciteBstWouldAddEndPuncttrue
\mciteSetBstMidEndSepPunct{\mcitedefaultmidpunct}
{\mcitedefaultendpunct}{\mcitedefaultseppunct}\relax
\EndOfBibitem
\bibitem[Ko \latin{et~al.}(2023)Ko, Calegari~Andrade, Sparrow, Zhang, and
  DiStasio]{SeA_RDJ}
Ko,~H.-Y.; Calegari~Andrade,~M.~F.; Sparrow,~Z.~M.; Zhang,~J.-a.; DiStasio,~R.
  A.~J. High-Throughput Condensed-Phase Hybrid Density Functional Theory for
  Large-Scale Finite-Gap Systems: The SeA Approach. \emph{J. Chem. Theory
  Comput.} \textbf{2023}, \emph{19}, 4182--4201\relax
\mciteBstWouldAddEndPuncttrue
\mciteSetBstMidEndSepPunct{\mcitedefaultmidpunct}
{\mcitedefaultendpunct}{\mcitedefaultseppunct}\relax
\EndOfBibitem
\bibitem[Guidon \latin{et~al.}(2008)Guidon, Schiffmann, Hutter, and
  VandeVondele]{HFX_Hutter_JCP}
Guidon,~M.; Schiffmann,~F.; Hutter,~J.; VandeVondele,~J. Ab initio molecular
  dynamics using hybrid density functionals. \emph{J. Chem. Phys.}
  \textbf{2008}, \emph{128}, 214104\relax
\mciteBstWouldAddEndPuncttrue
\mciteSetBstMidEndSepPunct{\mcitedefaultmidpunct}
{\mcitedefaultendpunct}{\mcitedefaultseppunct}\relax
\EndOfBibitem
\bibitem[Liberatore \latin{et~al.}(2018)Liberatore, Meli, and
  Rothlisberger]{MTS_AIMD_Ursula}
Liberatore,~E.; Meli,~R.; Rothlisberger,~U. A Versatile Multiple Time Step
  Scheme for Efficient ab Initio Molecular Dynamics Simulations.
  \emph{J.~Chem.~Theory Comput.} \textbf{2018}, \emph{14}, 2834--2842\relax
\mciteBstWouldAddEndPuncttrue
\mciteSetBstMidEndSepPunct{\mcitedefaultmidpunct}
{\mcitedefaultendpunct}{\mcitedefaultseppunct}\relax
\EndOfBibitem
\bibitem[Fatehi and Steele(2015)Fatehi, and Steele]{MTS_AIMD_Steele_3}
Fatehi,~S.; Steele,~R.~P. Multiple-Time Step Ab Initio Molecular Dynamics Based
  on Two-Electron Integral Screening. \emph{J.~Chem.~Theory Comput.}
  \textbf{2015}, \emph{11}, 884--898\relax
\mciteBstWouldAddEndPuncttrue
\mciteSetBstMidEndSepPunct{\mcitedefaultmidpunct}
{\mcitedefaultendpunct}{\mcitedefaultseppunct}\relax
\EndOfBibitem
\bibitem[Bircher and Rothlisberger(2018)Bircher, and
  Rothlisberger]{JPCL_2018_Bircher}
Bircher,~M.~P.; Rothlisberger,~U. Exploiting Coordinate Scaling Relations To
  Accelerate Exact Exchange Calculations. \emph{J. Phys. Chem. Lett.}
  \textbf{2018}, \emph{9}, 3886--3890\relax
\mciteBstWouldAddEndPuncttrue
\mciteSetBstMidEndSepPunct{\mcitedefaultmidpunct}
{\mcitedefaultendpunct}{\mcitedefaultseppunct}\relax
\EndOfBibitem
\bibitem[Bircher and Rothlisberger(2020)Bircher, and
  Rothlisberger]{CPC_BIRCHER_2020}
Bircher,~M.~P.; Rothlisberger,~U. From a week to less than a day: Speedup and
  scaling of coordinate-scaled exact exchange calculations in plane waves.
  \emph{Comput. Phys. Commun.} \textbf{2020}, \emph{247}, 106943\relax
\mciteBstWouldAddEndPuncttrue
\mciteSetBstMidEndSepPunct{\mcitedefaultmidpunct}
{\mcitedefaultendpunct}{\mcitedefaultseppunct}\relax
\EndOfBibitem
\bibitem[Weber \latin{et~al.}(2014)Weber, Bekas, Laino, Curioni, Bertsch, and
  Futral]{HFX_Curioni}
Weber,~V.; Bekas,~C.; Laino,~T.; Curioni,~A.; Bertsch,~A.; Futral,~S. Shedding
  Light on {L}ithium/Air Batteries Using Millions of Threads on the {BG/Q}
  Supercomputer. 2014 IEEE 28th International Parallel and Distributed
  Processing Symposium. Phoenix, AZ, USA, 2014; pp 735--744\relax
\mciteBstWouldAddEndPuncttrue
\mciteSetBstMidEndSepPunct{\mcitedefaultmidpunct}
{\mcitedefaultendpunct}{\mcitedefaultseppunct}\relax
\EndOfBibitem
\bibitem[Duchemin and Gygi(2010)Duchemin, and Gygi]{DUCHEMIN_2010}
Duchemin,~I.; Gygi,~F. A scalable and accurate algorithm for the computation of
  {H}artree–{F}ock exchange. \emph{Comput. Phys. Commun.} \textbf{2010},
  \emph{181}, 855 -- 860\relax
\mciteBstWouldAddEndPuncttrue
\mciteSetBstMidEndSepPunct{\mcitedefaultmidpunct}
{\mcitedefaultendpunct}{\mcitedefaultseppunct}\relax
\EndOfBibitem
\bibitem[Varini \latin{et~al.}(2013)Varini, Ceresoli, Martin-Samos, Girotto,
  and Cavazzoni]{VARINI_2013}
Varini,~N.; Ceresoli,~D.; Martin-Samos,~L.; Girotto,~I.; Cavazzoni,~C.
  Enhancement of {DFT}-calculations at petascale: Nuclear Magnetic Resonance,
  Hybrid Density Functional Theory and {C}ar–{P}arrinello calculations.
  \emph{Comput. Phys. Commun.} \textbf{2013}, \emph{184}, 1827--1833\relax
\mciteBstWouldAddEndPuncttrue
\mciteSetBstMidEndSepPunct{\mcitedefaultmidpunct}
{\mcitedefaultendpunct}{\mcitedefaultseppunct}\relax
\EndOfBibitem
\bibitem[Barnes \latin{et~al.}(2017)Barnes, Kurth, Carrier, Wichmann,
  Prendergast, Kent, and Deslippe]{BARNES_2017}
Barnes,~T.~A.; Kurth,~T.; Carrier,~P.; Wichmann,~N.; Prendergast,~D.;
  Kent,~P.~R.; Deslippe,~J. Improved treatment of exact exchange in {Q}uantum
  {ESPRESSO}. \emph{Comput. Phys. Commun.} \textbf{2017}, \emph{214},
  52--58\relax
\mciteBstWouldAddEndPuncttrue
\mciteSetBstMidEndSepPunct{\mcitedefaultmidpunct}
{\mcitedefaultendpunct}{\mcitedefaultseppunct}\relax
\EndOfBibitem
\bibitem[Mandal \latin{et~al.}(2022)Mandal, Kar, Klöffel, Meyer, and
  Nair]{sagar_JCC_scaling}
Mandal,~S.; Kar,~R.; Klöffel,~T.; Meyer,~B.; Nair,~N.~N. Improving the scaling
  and performance of multiple time stepping-based molecular dynamics with
  hybrid density functionals. \emph{J. Comput. Chem.} \textbf{2022}, \emph{43},
  588--597\relax
\mciteBstWouldAddEndPuncttrue
\mciteSetBstMidEndSepPunct{\mcitedefaultmidpunct}
{\mcitedefaultendpunct}{\mcitedefaultseppunct}\relax
\EndOfBibitem
\bibitem[Bolnykh \latin{et~al.}(2019)Bolnykh, Olsen, Meloni, Bircher, Ippoliti,
  Carloni, and Rothlisberger]{jctc_Bolnykh_2019}
Bolnykh,~V.; Olsen,~J. M.~H.; Meloni,~S.; Bircher,~M.~P.; Ippoliti,~E.;
  Carloni,~P.; Rothlisberger,~U. Extreme Scalability of {DFT}-Based {QM/MM MD}
  Simulations Using {MiMiC}. \emph{J. Chem. Theory Comput.} \textbf{2019},
  \emph{15}, 5601--5613\relax
\mciteBstWouldAddEndPuncttrue
\mciteSetBstMidEndSepPunct{\mcitedefaultmidpunct}
{\mcitedefaultendpunct}{\mcitedefaultseppunct}\relax
\EndOfBibitem
\bibitem[Vinson(2020)]{single_precision_hfx}
Vinson,~J. Faster exact exchange in periodic systems using single-precision
  arithmetic. \emph{J. Chem. Phys.} \textbf{2020}, \emph{153}, 204106\relax
\mciteBstWouldAddEndPuncttrue
\mciteSetBstMidEndSepPunct{\mcitedefaultmidpunct}
{\mcitedefaultendpunct}{\mcitedefaultseppunct}\relax
\EndOfBibitem
\bibitem[von Rudorff \latin{et~al.}(2017)von Rudorff, Jakobsen, Rosso, and
  Blumberger]{HFX_JB}
von Rudorff,~G.~F.; Jakobsen,~R.; Rosso,~K.~M.; Blumberger,~J. Improving the
  Performance of Hybrid Functional-Based Molecular Dynamics Simulation through
  Screening of Hartree–Fock Exchange Forces. \emph{J.~Chem.~Theory Comput.}
  \textbf{2017}, \emph{13}, 2178--2184\relax
\mciteBstWouldAddEndPuncttrue
\mciteSetBstMidEndSepPunct{\mcitedefaultmidpunct}
{\mcitedefaultendpunct}{\mcitedefaultseppunct}\relax
\EndOfBibitem
\bibitem[Ratcliff \latin{et~al.}(2018)Ratcliff, Degomme, Flores-Livas,
  Goedecker, and Genovese]{HFX_Goedecker}
Ratcliff,~L.~E.; Degomme,~A.; Flores-Livas,~J.~A.; Goedecker,~S.; Genovese,~L.
  Affordable and accurate large-scale hybrid-functional calculations on
  \uppercase{GPU}-accelerated supercomputers. \emph{J. Phys.: Condens. Matter}
  \textbf{2018}, \emph{30}, 095901\relax
\mciteBstWouldAddEndPuncttrue
\mciteSetBstMidEndSepPunct{\mcitedefaultmidpunct}
{\mcitedefaultendpunct}{\mcitedefaultseppunct}\relax
\EndOfBibitem
\bibitem[Lin(2016)]{ACE_Lin}
Lin,~L. Adaptively Compressed Exchange Operator. \emph{J.~Chem.~Theory Comput.}
  \textbf{2016}, \emph{12}, 2242--2249\relax
\mciteBstWouldAddEndPuncttrue
\mciteSetBstMidEndSepPunct{\mcitedefaultmidpunct}
{\mcitedefaultendpunct}{\mcitedefaultseppunct}\relax
\EndOfBibitem
\bibitem[Hu \latin{et~al.}(2017)Hu, Lin, Banerjee, Vecharynski, and
  Yang]{ACE_Lin_1}
Hu,~W.; Lin,~L.; Banerjee,~A.~S.; Vecharynski,~E.; Yang,~C. Adaptively
  Compressed Exchange Operator for Large-Scale Hybrid Density Functional
  Calculations with Applications to the Adsorption of Water on Silicene.
  \emph{J.~Chem.~Theory Comput.} \textbf{2017}, \emph{13}, 1188--1198\relax
\mciteBstWouldAddEndPuncttrue
\mciteSetBstMidEndSepPunct{\mcitedefaultmidpunct}
{\mcitedefaultendpunct}{\mcitedefaultseppunct}\relax
\EndOfBibitem
\bibitem[Chen \latin{et~al.}(2023)Chen, Wu, Hu, and Yang]{ACE_2023}
Chen,~S.; Wu,~K.; Hu,~W.; Yang,~J. Low-rank approximations for accelerating
  plane-wave hybrid functional calculations in unrestricted and noncollinear
  spin density functional theory. \emph{J. Chem. Phys.} \textbf{2023},
  \emph{158}, 134106\relax
\mciteBstWouldAddEndPuncttrue
\mciteSetBstMidEndSepPunct{\mcitedefaultmidpunct}
{\mcitedefaultendpunct}{\mcitedefaultseppunct}\relax
\EndOfBibitem
\bibitem[Mandal and Nair(2019)Mandal, and Nair]{JCP_2019_sagar}
Mandal,~S.; Nair,~N.~N. Speeding-up ab initio molecular dynamics with hybrid
  functionals using adaptively compressed exchange operator based multiple
  timestepping. \emph{J. Chem. Phys.} \textbf{2019}, \emph{151}, 151102\relax
\mciteBstWouldAddEndPuncttrue
\mciteSetBstMidEndSepPunct{\mcitedefaultmidpunct}
{\mcitedefaultendpunct}{\mcitedefaultseppunct}\relax
\EndOfBibitem
\bibitem[Mandal and Nair(2020)Mandal, and Nair]{sagar_JCC}
Mandal,~S.; Nair,~N.~N. Efficient computation of free energy surfaces of
  chemical reactions using ab initio molecular dynamics with hybrid functionals
  and plane waves. \emph{J. Comput. Chem.} \textbf{2020}, \emph{41},
  1790--1797\relax
\mciteBstWouldAddEndPuncttrue
\mciteSetBstMidEndSepPunct{\mcitedefaultmidpunct}
{\mcitedefaultendpunct}{\mcitedefaultseppunct}\relax
\EndOfBibitem
\bibitem[Tuckerman \latin{et~al.}(1992)Tuckerman, Berne, and Martyna]{respa}
Tuckerman,~M.; Berne,~B.~J.; Martyna,~G.~J. {Reversible multiple time scale
  molecular dynamics}. \emph{J. Chem. Phys.} \textbf{1992}, \emph{97},
  1990--2001\relax
\mciteBstWouldAddEndPuncttrue
\mciteSetBstMidEndSepPunct{\mcitedefaultmidpunct}
{\mcitedefaultendpunct}{\mcitedefaultseppunct}\relax
\EndOfBibitem
\bibitem[Damle \latin{et~al.}(2015)Damle, Lin, and Ying]{SCDM_main}
Damle,~A.; Lin,~L.; Ying,~L. Compressed Representation of {K}ohn–{S}ham
  Orbitals via Selected Columns of the Density Matrix. \emph{J. Chem. Theory
  Comput.} \textbf{2015}, \emph{11}, 1463--1469\relax
\mciteBstWouldAddEndPuncttrue
\mciteSetBstMidEndSepPunct{\mcitedefaultmidpunct}
{\mcitedefaultendpunct}{\mcitedefaultseppunct}\relax
\EndOfBibitem
\bibitem[Mandal \latin{et~al.}(2021)Mandal, Thakkur, and Nair]{sagar_JCTC}
Mandal,~S.; Thakkur,~V.; Nair,~N.~N. Achieving an Order of Magnitude Speedup in
  Hybrid-Functional- and Plane-Wave-Based Ab Initio Molecular Dynamics:
  Applications to Proton-Transfer Reactions in Enzymes and in Solution.
  \emph{J.~Chem.~Theory Comput.} \textbf{2021}, \emph{17}, 2244--2255\relax
\mciteBstWouldAddEndPuncttrue
\mciteSetBstMidEndSepPunct{\mcitedefaultmidpunct}
{\mcitedefaultendpunct}{\mcitedefaultseppunct}\relax
\EndOfBibitem
\bibitem[Klöffel \latin{et~al.}(2021)Klöffel, Mathias, and
  Meyer]{KLOFFEL2021}
Klöffel,~T.; Mathias,~G.; Meyer,~B. Integrating state of the art compute,
  communication, and autotuning strategies to multiply the performance of ab
  initio molecular dynamics on massively parallel multi-core supercomputers.
  \emph{Comput. Phys. Commun.} \textbf{2021}, \emph{260}, 107745\relax
\mciteBstWouldAddEndPuncttrue
\mciteSetBstMidEndSepPunct{\mcitedefaultmidpunct}
{\mcitedefaultendpunct}{\mcitedefaultseppunct}\relax
\EndOfBibitem
\bibitem[Schlick \latin{et~al.}(1998)Schlick, Mandziuk, Skeel, and
  Srinivas]{Resonace_1}
Schlick,~T.; Mandziuk,~M.; Skeel,~R.~D.; Srinivas,~K. Nonlinear Resonance
  Artifacts in Molecular Dynamics Simulations. \emph{J. Comput. Phys.}
  \textbf{1998}, \emph{140}, 1--29\relax
\mciteBstWouldAddEndPuncttrue
\mciteSetBstMidEndSepPunct{\mcitedefaultmidpunct}
{\mcitedefaultendpunct}{\mcitedefaultseppunct}\relax
\EndOfBibitem
\bibitem[Sandu and Schlick(1999)Sandu, and Schlick]{Resonace_2}
Sandu,~A.; Schlick,~T. Masking resonance artifacts in force-splitting methods
  for biomolecular simulations by extrapolative Langevin dynamics. \emph{J.
  Comput. Phys.} \textbf{1999}, \emph{151}, 74--113\relax
\mciteBstWouldAddEndPuncttrue
\mciteSetBstMidEndSepPunct{\mcitedefaultmidpunct}
{\mcitedefaultendpunct}{\mcitedefaultseppunct}\relax
\EndOfBibitem
\bibitem[Ma \latin{et~al.}(2003)Ma, Izaguirre, and Skeel]{Resonace_3}
Ma,~Q.; Izaguirre,~J.~A.; Skeel,~R.~D. Verlet-I/R-RESPA/Impulse is Limited by
  Nonlinear Instabilities. \emph{SIAM J. Sci. Comput.} \textbf{2003},
  \emph{24}, 1951--1973\relax
\mciteBstWouldAddEndPuncttrue
\mciteSetBstMidEndSepPunct{\mcitedefaultmidpunct}
{\mcitedefaultendpunct}{\mcitedefaultseppunct}\relax
\EndOfBibitem
\bibitem[Izaguirre \latin{et~al.}(1999)Izaguirre, Reich, and
  Skeel]{Resonace_free_1}
Izaguirre,~J.~A.; Reich,~S.; Skeel,~R.~D. Longer time steps for molecular
  dynamics. \emph{J. Chem. Phys.} \textbf{1999}, \emph{110}, 9853--9864\relax
\mciteBstWouldAddEndPuncttrue
\mciteSetBstMidEndSepPunct{\mcitedefaultmidpunct}
{\mcitedefaultendpunct}{\mcitedefaultseppunct}\relax
\EndOfBibitem
\bibitem[Izaguirre \latin{et~al.}(2001)Izaguirre, Catarello, Wozniak, , and
  Skeel]{Resonace_free_2}
Izaguirre,~J.~A.; Catarello,~D.~P.; Wozniak,~J.~M.; ; Skeel,~R.~D. Langevin
  stabilization of molecular dynamics. \emph{J. Chem. Phys.} \textbf{2001},
  \emph{114}, 2090--2098\relax
\mciteBstWouldAddEndPuncttrue
\mciteSetBstMidEndSepPunct{\mcitedefaultmidpunct}
{\mcitedefaultendpunct}{\mcitedefaultseppunct}\relax
\EndOfBibitem
\bibitem[Minary \latin{et~al.}(2004)Minary, Tuckerman, and
  Martyna]{Resonace_free_3}
Minary,~P.; Tuckerman,~M.~E.; Martyna,~G.~J. Long Time Molecular Dynamics for
  Enhanced Conformational Sampling in Biomolecular Systems. \emph{Phys. Rev.
  Lett.} \textbf{2004}, \emph{93}, 150201\relax
\mciteBstWouldAddEndPuncttrue
\mciteSetBstMidEndSepPunct{\mcitedefaultmidpunct}
{\mcitedefaultendpunct}{\mcitedefaultseppunct}\relax
\EndOfBibitem
\bibitem[Omelyan and Kovalenko(2011)Omelyan, and Kovalenko]{Resonace_free_4}
Omelyan,~I.~P.; Kovalenko,~A. Multiple time scale molecular dynamics for fluids
  with orientational degrees of freedom. II. Canonical and isokinetic
  ensembles. \emph{J. Chem. Phys.} \textbf{2011}, \emph{135}, 234107\relax
\mciteBstWouldAddEndPuncttrue
\mciteSetBstMidEndSepPunct{\mcitedefaultmidpunct}
{\mcitedefaultendpunct}{\mcitedefaultseppunct}\relax
\EndOfBibitem
\bibitem[Omelyan and Kovalenko(2012)Omelyan, and Kovalenko]{Resonace_free_5}
Omelyan,~I.~P.; Kovalenko,~A. Overcoming the Barrier on Time Step Size in
  Multiscale Molecular Dynamics Simulation of Molecular Liquids. \emph{J. Chem.
  Theory Comput.} \textbf{2012}, \emph{8}, 6--16\relax
\mciteBstWouldAddEndPuncttrue
\mciteSetBstMidEndSepPunct{\mcitedefaultmidpunct}
{\mcitedefaultendpunct}{\mcitedefaultseppunct}\relax
\EndOfBibitem
\bibitem[Omelyan and Kovalenko(2011)Omelyan, and Kovalenko]{Resonace_free_6}
Omelyan,~I.~P.; Kovalenko,~A. Efficient multiple time scale molecular dynamics:
  Using colored noise thermostats to stabilize resonances. \emph{J. Chem.Phys.}
  \textbf{2011}, \emph{134}, 014103\relax
\mciteBstWouldAddEndPuncttrue
\mciteSetBstMidEndSepPunct{\mcitedefaultmidpunct}
{\mcitedefaultendpunct}{\mcitedefaultseppunct}\relax
\EndOfBibitem
\bibitem[Leimkuhler \latin{et~al.}(2013)Leimkuhler, Margul, and
  Tuckerman]{SINR_main}
Leimkuhler,~B.; Margul,~D.~T.; Tuckerman,~M.~E. Stochastic, resonance-free
  multiple time-step algorithm for molecular dynamics with very large time
  steps. \emph{Mol. Phys.} \textbf{2013}, \emph{111}, 3579--3594\relax
\mciteBstWouldAddEndPuncttrue
\mciteSetBstMidEndSepPunct{\mcitedefaultmidpunct}
{\mcitedefaultendpunct}{\mcitedefaultseppunct}\relax
\EndOfBibitem
\bibitem[Margul and Tuckerman(2016)Margul, and Tuckerman]{SINR_polarised}
Margul,~D.~T.; Tuckerman,~M.~E. A Stochastic, Resonance-Free Multiple Time-Step
  Algorithm for Polarizable Models That Permits Very Large Time Steps. \emph{J.
  Chem. Theory Comput.} \textbf{2016}, \emph{12}, 2170--2180\relax
\mciteBstWouldAddEndPuncttrue
\mciteSetBstMidEndSepPunct{\mcitedefaultmidpunct}
{\mcitedefaultendpunct}{\mcitedefaultseppunct}\relax
\EndOfBibitem
\bibitem[Zhang \latin{et~al.}(2019)Zhang, Liu, Yan, Tuckerman, and
  Liu]{SINR-middle}
Zhang,~Z.; Liu,~X.; Yan,~K.; Tuckerman,~M.~E.; Liu,~J. Unified Efficient
  Thermostat Scheme for the Canonical Ensemble with Holonomic or Isokinetic
  Constraints via Molecular Dynamics. \emph{J. Phys. Chem. A} \textbf{2019},
  \emph{123}, 6056--6079\relax
\mciteBstWouldAddEndPuncttrue
\mciteSetBstMidEndSepPunct{\mcitedefaultmidpunct}
{\mcitedefaultendpunct}{\mcitedefaultseppunct}\relax
\EndOfBibitem
\bibitem[Abreu and Tuckerman(2021)Abreu, and Tuckerman]{SINR_MP_21}
Abreu,~C. R.~A.; Tuckerman,~M.~E. Hamiltonian based resonance-free approach for
  enabling very large time steps in multiple time-scale molecular dynamics.
  \emph{Mol. Phys.} \textbf{2021}, \emph{119}, e1923848\relax
\mciteBstWouldAddEndPuncttrue
\mciteSetBstMidEndSepPunct{\mcitedefaultmidpunct}
{\mcitedefaultendpunct}{\mcitedefaultseppunct}\relax
\EndOfBibitem
\bibitem[Abreu and Tuckerman(2020)Abreu, and Tuckerman]{SINR_20_jctc}
Abreu,~C.~R.; Tuckerman,~M.~E. Molecular Dynamics with Very Large Time Steps
  for the Calculation of Solvation Free Energies. \emph{J. Chem. Theory
  Comput.} \textbf{2020}, \emph{16}, 7314--7327\relax
\mciteBstWouldAddEndPuncttrue
\mciteSetBstMidEndSepPunct{\mcitedefaultmidpunct}
{\mcitedefaultendpunct}{\mcitedefaultseppunct}\relax
\EndOfBibitem
\bibitem[Abreu and Tuckerman(2021)Abreu, and Tuckerman]{SINR_21_epj}
Abreu,~C.; Tuckerman,~M. Multiple timescale molecular dynamics with very large
  time steps: avoidance of resonances. \emph{Eur. Phys. J. B} \textbf{2021},
  \emph{94}, 231\relax
\mciteBstWouldAddEndPuncttrue
\mciteSetBstMidEndSepPunct{\mcitedefaultmidpunct}
{\mcitedefaultendpunct}{\mcitedefaultseppunct}\relax
\EndOfBibitem
\bibitem[{Hutter et al.}()]{cpmd}
{Hutter et al.},~J. {CPMD}: {A}n {A}b {I}nitio {E}lectronic {S}tructure and
  {M}olecular {D}ynamics {P}rogram. see {\tt http://www.cpmd.org} (accessed on
  December 31, 2020)\relax
\mciteBstWouldAddEndPuncttrue
\mciteSetBstMidEndSepPunct{\mcitedefaultmidpunct}
{\mcitedefaultendpunct}{\mcitedefaultseppunct}\relax
\EndOfBibitem
\bibitem[Hutter and Curioni(2005)Hutter, and Curioni]{CPG_curioni}
Hutter,~J.; Curioni,~A. Car–Parrinello Molecular Dynamics on Massively
  Parallel Computers. \emph{ChemPhysChem} \textbf{2005}, \emph{6},
  1788--1793\relax
\mciteBstWouldAddEndPuncttrue
\mciteSetBstMidEndSepPunct{\mcitedefaultmidpunct}
{\mcitedefaultendpunct}{\mcitedefaultseppunct}\relax
\EndOfBibitem
\bibitem[Evans and Morriss(1984)Evans, and Morriss]{Isokinetic_1}
Evans,~D.~J.; Morriss,~G.~P. Non-Newtonian molecular dynamics. \emph{Comput.
  Phys. Rep.} \textbf{1984}, \emph{1}, 299--343\relax
\mciteBstWouldAddEndPuncttrue
\mciteSetBstMidEndSepPunct{\mcitedefaultmidpunct}
{\mcitedefaultendpunct}{\mcitedefaultseppunct}\relax
\EndOfBibitem
\bibitem[Evans and Morriss(1990)Evans, and Morriss]{Evans-Morriss-book}
Evans,~D.~J.; Morriss,~G.~P. \emph{Statistical Mechanics of Nonequilibrium
  Liquids}; Academic: London, 1990\relax
\mciteBstWouldAddEndPuncttrue
\mciteSetBstMidEndSepPunct{\mcitedefaultmidpunct}
{\mcitedefaultendpunct}{\mcitedefaultseppunct}\relax
\EndOfBibitem
\bibitem[Minary \latin{et~al.}(2003)Minary, Martyna, and
  Tuckerman]{Isokinetic_3}
Minary,~P.; Martyna,~G.~J.; Tuckerman,~M.~E. Algorithms and novel applications
  based on the isokinetic ensemble. I. Biophysical and path integral molecular
  dynamics. \emph{J. Chem. Phys.} \textbf{2003}, \emph{118}, 2510--2526\relax
\mciteBstWouldAddEndPuncttrue
\mciteSetBstMidEndSepPunct{\mcitedefaultmidpunct}
{\mcitedefaultendpunct}{\mcitedefaultseppunct}\relax
\EndOfBibitem
\bibitem[Troullier and Martins(1991)Troullier, and Martins]{PRB_TM}
Troullier,~N.; Martins,~J.~L. Efficient pseudopotentials for plane-wave
  calculations. \emph{Phys. Rev. B} \textbf{1991}, \emph{43}, 1993--2006\relax
\mciteBstWouldAddEndPuncttrue
\mciteSetBstMidEndSepPunct{\mcitedefaultmidpunct}
{\mcitedefaultendpunct}{\mcitedefaultseppunct}\relax
\EndOfBibitem
\bibitem[Pulay(1980)]{PULAY_DIIS}
Pulay,~P. Convergence acceleration of iterative sequences. the case of
  \uppercase{scf} iteration. \emph{Chem. Phys. Lett.} \textbf{1980}, \emph{73},
  393 -- 398\relax
\mciteBstWouldAddEndPuncttrue
\mciteSetBstMidEndSepPunct{\mcitedefaultmidpunct}
{\mcitedefaultendpunct}{\mcitedefaultseppunct}\relax
\EndOfBibitem
\bibitem[Hutter \latin{et~al.}(1994)Hutter, Lüthi, and
  Parrinello]{HUTTER_DIIS}
Hutter,~J.; Lüthi,~H.~P.; Parrinello,~M. Electronic structure optimization in
  plane-wave-based density functional calculations by direct inversion in the
  iterative subspace. \emph{Comput. Mater. Sci.} \textbf{1994}, \emph{2}, 244
  -- 248\relax
\mciteBstWouldAddEndPuncttrue
\mciteSetBstMidEndSepPunct{\mcitedefaultmidpunct}
{\mcitedefaultendpunct}{\mcitedefaultseppunct}\relax
\EndOfBibitem
\bibitem[\ifmmode~\check{S}\else \v{S}\fi{}tich
  \latin{et~al.}(1989)\ifmmode~\check{S}\else \v{S}\fi{}tich, Car, Parrinello,
  and Baroni]{PRB_PCG}
\ifmmode~\check{S}\else \v{S}\fi{}tich,~I.; Car,~R.; Parrinello,~M.; Baroni,~S.
  Conjugate gradient minimization of the energy functional: A new method for
  electronic structure calculation. \emph{Phys. Rev. B} \textbf{1989},
  \emph{39}, 4997--5004\relax
\mciteBstWouldAddEndPuncttrue
\mciteSetBstMidEndSepPunct{\mcitedefaultmidpunct}
{\mcitedefaultendpunct}{\mcitedefaultseppunct}\relax
\EndOfBibitem
\bibitem[Kolafa(2004)]{JCC_ASPC}
Kolafa,~J. Time-reversible always stable predictor–corrector method for
  molecular dynamics of polarizable molecules. \emph{J. Comput. Chem.}
  \textbf{2004}, \emph{25}, 335--342\relax
\mciteBstWouldAddEndPuncttrue
\mciteSetBstMidEndSepPunct{\mcitedefaultmidpunct}
{\mcitedefaultendpunct}{\mcitedefaultseppunct}\relax
\EndOfBibitem
\bibitem[VandeVondele \latin{et~al.}(2006)VandeVondele, Sulpizi, and
  Sprik]{BQ-ang}
VandeVondele,~J.; Sulpizi,~M.; Sprik,~M. From Solvent Fluctuations to
  Quantitative Redox Properties of Quinones in Methanol and Acetonitrile.
  \emph{Angew. Chem. Int. Ed.} \textbf{2006}, \emph{45}, 1936--1938\relax
\mciteBstWouldAddEndPuncttrue
\mciteSetBstMidEndSepPunct{\mcitedefaultmidpunct}
{\mcitedefaultendpunct}{\mcitedefaultseppunct}\relax
\EndOfBibitem
\bibitem[Mandal \latin{et~al.}(2023)Mandal, Kar, Meyer, and Nair]{sagar_Fe_CPC}
Mandal,~S.; Kar,~R.; Meyer,~B.; Nair,~N.~N. Hybrid Functional and Plane Waves
  based Ab Initio Molecular Dynamics Study of the Aqueous Fe2+/Fe3+ Redox
  Reaction. \emph{ChemPhysChem} \textbf{2023}, \emph{24}, e202200617\relax
\mciteBstWouldAddEndPuncttrue
\mciteSetBstMidEndSepPunct{\mcitedefaultmidpunct}
{\mcitedefaultendpunct}{\mcitedefaultseppunct}\relax
\EndOfBibitem
\bibitem[Kowalski \latin{et~al.}(2009)Kowalski, Meyer, and
  Marx]{Marx_PRB_static_tio2}
Kowalski,~P.~M.; Meyer,~B.; Marx,~D. Composition, structure, and stability of
  the rutile ${\text{TiO}}_{2}(110)$ surface: Oxygen depletion, hydroxylation,
  hydrogen migration, and water adsorption. \emph{Phys. Rev. B} \textbf{2009},
  \emph{79}, 115410\relax
\mciteBstWouldAddEndPuncttrue
\mciteSetBstMidEndSepPunct{\mcitedefaultmidpunct}
{\mcitedefaultendpunct}{\mcitedefaultseppunct}\relax
\EndOfBibitem
\bibitem[Tripathi and Nair(2016)Tripathi, and Nair]{Ravi_2016_JPCB}
Tripathi,~R.; Nair,~N.~N. Deacylation Mechanism and Kinetics of Acyl--Enzyme
  Complex of Class--C $\beta$--Lactamase and Cephalothin. \emph{J. Phys. Chem.
  B} \textbf{2016}, \emph{120}, 2681--2690\relax
\mciteBstWouldAddEndPuncttrue
\mciteSetBstMidEndSepPunct{\mcitedefaultmidpunct}
{\mcitedefaultendpunct}{\mcitedefaultseppunct}\relax
\EndOfBibitem
\bibitem[Cheatham~III \latin{et~al.}(1999)Cheatham~III, Cieplak, and
  Kollman]{parm99}
Cheatham~III,~T.~E.; Cieplak,~P.; Kollman,~P.~A. A Modified Version of the
  {C}ornell et al. Force Field with Improved Sugar Pucker Phases and Helical
  Repeat. \emph{J. Biomol. Struct. Dyn.} \textbf{1999}, \emph{16},
  845--862\relax
\mciteBstWouldAddEndPuncttrue
\mciteSetBstMidEndSepPunct{\mcitedefaultmidpunct}
{\mcitedefaultendpunct}{\mcitedefaultseppunct}\relax
\EndOfBibitem
\bibitem[Woods and Chappelle(2000)Woods, and Chappelle]{RESP_2000}
Woods,~R.; Chappelle,~R. Restrained electrostatic potential atomic partial
  charges for condensed-phase simulations of carbohydrates. \emph{J. Mol.
  Struct.: THEOCHEM} \textbf{2000}, \emph{527}, 149--156\relax
\mciteBstWouldAddEndPuncttrue
\mciteSetBstMidEndSepPunct{\mcitedefaultmidpunct}
{\mcitedefaultendpunct}{\mcitedefaultseppunct}\relax
\EndOfBibitem
\bibitem[Wang \latin{et~al.}(2004)Wang, Wolf, Caldwell, Kollman, and
  Case]{gaff}
Wang,~J.; Wolf,~R.~M.; Caldwell,~J.~W.; Kollman,~P.~A.; Case,~D.~A. Development
  and testing of a general amber force field. \emph{J. Comput. Chem.}
  \textbf{2004}, \emph{25}, 1157--1174\relax
\mciteBstWouldAddEndPuncttrue
\mciteSetBstMidEndSepPunct{\mcitedefaultmidpunct}
{\mcitedefaultendpunct}{\mcitedefaultseppunct}\relax
\EndOfBibitem
\bibitem[Case \latin{et~al.}(2018)Case, Ben-Shalom, Brozell, Cerutti, III,
  Cruzeiro, T.A.~Darden, Ghoreishi, Gilson, H.~Gohlke, Harris, Homeyer, Huang,
  Izadi, Kovalenko, Kurtzman, Lee, LeGrand, Li, Lin, Liu, Luchko, Luo,
  Mermelstein, Merz, Miao, Monard, Nguyen, Nguyen, Omelyan, Onufriev, Pan, Qi,
  Roe, Roitberg, Sagui, Schott-Verdugo, Shen, Simmerling, Smith, SalomonFerrer,
  Swails, Walker, Wang, Wei, Wolf, Wu, Xiao, York, and Kollman]{amber18}
Case,~D.; Ben-Shalom,~I.; Brozell,~S.; Cerutti,~D.; III,~T.~C.; Cruzeiro,~V.;
  T.A.~Darden,~R.~D.; Ghoreishi,~D.; Gilson,~M.; H.~Gohlke,~D.~G.,~A.W.~Goetz;
  Harris,~R.; Homeyer,~N.; Huang,~Y.; Izadi,~S.; Kovalenko,~A.; Kurtzman,~T.;
  Lee,~T.; LeGrand,~S.; Li,~P.; Lin,~C.; Liu,~J.; Luchko,~T.; Luo,~R.;
  Mermelstein,~D.; Merz,~K.; Miao,~Y.; Monard,~G.; Nguyen,~C.; Nguyen,~H.;
  Omelyan,~I.; Onufriev,~A.; Pan,~F.; Qi,~R.; Roe,~D.; Roitberg,~A.; Sagui,~C.;
  Schott-Verdugo,~S.; Shen,~J.; Simmerling,~C.; Smith,~J.; SalomonFerrer,~R.;
  Swails,~J.; Walker,~R.; Wang,~J.; Wei,~H.; Wolf,~R.; Wu,~X.; Xiao,~L.;
  York,~D.; Kollman,~P. \emph{Amber 2018}; 2018\relax
\mciteBstWouldAddEndPuncttrue
\mciteSetBstMidEndSepPunct{\mcitedefaultmidpunct}
{\mcitedefaultendpunct}{\mcitedefaultseppunct}\relax
\EndOfBibitem
\bibitem[Laio \latin{et~al.}(2002)Laio, VandeVondele, and
  Rothlisberger]{Laio_2002_JCP}
Laio,~A.; VandeVondele,~J.; Rothlisberger,~U. A Hamiltonian electrostatic
  coupling scheme for hybrid Car-Parrinello molecular dynamics simulations.
  \emph{J. Chem. Phys.} \textbf{2002}, \emph{116}, 6941--6947\relax
\mciteBstWouldAddEndPuncttrue
\mciteSetBstMidEndSepPunct{\mcitedefaultmidpunct}
{\mcitedefaultendpunct}{\mcitedefaultseppunct}\relax
\EndOfBibitem
\bibitem[Laio \latin{et~al.}(2002)Laio, VandeVondele, and
  Rothlisberger]{D-RESP}
Laio,~A.; VandeVondele,~J.; Rothlisberger,~U. {D-RESP: Dynamically Generated
  Electrostatic Potential Derived Charges from Quantum Mechanics/Molecular
  Mechanics Simulations}. \emph{J. Phys. Chem. B} \textbf{2002}, \emph{106},
  7300--7307\relax
\mciteBstWouldAddEndPuncttrue
\mciteSetBstMidEndSepPunct{\mcitedefaultmidpunct}
{\mcitedefaultendpunct}{\mcitedefaultseppunct}\relax
\EndOfBibitem
\bibitem[Scott \latin{et~al.}(1999)Scott, Hünenberger, Tironi, Mark, Billeter,
  Fennen, Torda, Huber, Krüger, and van Gunsteren]{Gromos96_package}
Scott,~W. R.~P.; Hünenberger,~P.~H.; Tironi,~I.~G.; Mark,~A.~E.;
  Billeter,~S.~R.; Fennen,~J.; Torda,~A.~E.; Huber,~T.; Krüger,~P.; van
  Gunsteren,~W.~F. The GROMOS Biomolecular Simulation Program Package. \emph{J.
  Phys. Chem. A} \textbf{1999}, \emph{103}, 3596--3607\relax
\mciteBstWouldAddEndPuncttrue
\mciteSetBstMidEndSepPunct{\mcitedefaultmidpunct}
{\mcitedefaultendpunct}{\mcitedefaultseppunct}\relax
\EndOfBibitem
\bibitem[Gro()]{Gromos96_note}
{Modified GROMOS96 version as distributed with the QM/MM Interface developed
  for CPMD.}\relax
\mciteBstWouldAddEndPunctfalse
\mciteSetBstMidEndSepPunct{\mcitedefaultmidpunct}
{}{\mcitedefaultseppunct}\relax
\EndOfBibitem
\bibitem[Awasthi \latin{et~al.}(2016)Awasthi, Kapil, and Nair]{JCC_shalini}
Awasthi,~S.; Kapil,~V.; Nair,~N.~N. Sampling free energy surfaces as slices by
  combining umbrella sampling and metadynamics. \emph{J. Comput. Chem.}
  \textbf{2016}, \emph{37}, 1413--1424\relax
\mciteBstWouldAddEndPuncttrue
\mciteSetBstMidEndSepPunct{\mcitedefaultmidpunct}
{\mcitedefaultendpunct}{\mcitedefaultseppunct}\relax
\EndOfBibitem
\bibitem[Torrie and Valleau(1974)Torrie, and Valleau]{US_method}
Torrie,~G.~M.; Valleau,~J.~P. Monte {C}arlo free energy estimates using
  non-{B}oltzmann sampling: Application to the sub-critical
  \uppercase{L}ennard-\uppercase{J}ones fluid. \emph{Chem. Phys. Lett.}
  \textbf{1974}, \emph{28}, 578 -- 581\relax
\mciteBstWouldAddEndPuncttrue
\mciteSetBstMidEndSepPunct{\mcitedefaultmidpunct}
{\mcitedefaultendpunct}{\mcitedefaultseppunct}\relax
\EndOfBibitem
\bibitem[Barducci \latin{et~al.}(2008)Barducci, Bussi, and Parrinello]{WT-MTD}
Barducci,~A.; Bussi,~G.; Parrinello,~M. Well-Tempered Metadynamics: A Smoothly
  Converging and Tunable Free-Energy Method. \emph{Phys. Rev. Lett.}
  \textbf{2008}, \emph{100}, 020603\relax
\mciteBstWouldAddEndPuncttrue
\mciteSetBstMidEndSepPunct{\mcitedefaultmidpunct}
{\mcitedefaultendpunct}{\mcitedefaultseppunct}\relax
\EndOfBibitem
\bibitem[Son \latin{et~al.}(2016)Son, Kim, Kim, and Park]{quinone_2016}
Son,~E.~J.; Kim,~J.~H.; Kim,~K.; Park,~C.~B. Quinone and its derivatives for
  energy harvesting and storage materials. \emph{J. Mater. Chem. A}
  \textbf{2016}, \emph{4}, 11179--11202\relax
\mciteBstWouldAddEndPuncttrue
\mciteSetBstMidEndSepPunct{\mcitedefaultmidpunct}
{\mcitedefaultendpunct}{\mcitedefaultseppunct}\relax
\EndOfBibitem
\bibitem[Dahlin \latin{et~al.}(1984)Dahlin, Miwa, Lu, and Nelson]{quinone_PNAS}
Dahlin,~D.~C.; Miwa,~G.~T.; Lu,~A.~Y.; Nelson,~S.~D. N-acetyl-p-benzoquinone
  imine: a cytochrome P-450-mediated oxidation product of acetaminophen.
  \emph{Proc. Natl. Acad. Sci.} \textbf{1984}, \emph{81}, 1327--1331\relax
\mciteBstWouldAddEndPuncttrue
\mciteSetBstMidEndSepPunct{\mcitedefaultmidpunct}
{\mcitedefaultendpunct}{\mcitedefaultseppunct}\relax
\EndOfBibitem
\bibitem[Kepler \latin{et~al.}(2019)Kepler, Zeller, and Rosokha]{quinone_JACS}
Kepler,~S.; Zeller,~M.; Rosokha,~S.~V. Anion--$\pi$ Complexes of Halides with
  p-Benzoquinones: Structures, Thermodynamics, and Criteria of Charge Transfer
  to Electron Transfer Transition. \emph{J. Am. Chem. Soc.} \textbf{2019},
  \emph{141}, 9338--9348\relax
\mciteBstWouldAddEndPuncttrue
\mciteSetBstMidEndSepPunct{\mcitedefaultmidpunct}
{\mcitedefaultendpunct}{\mcitedefaultseppunct}\relax
\EndOfBibitem
\bibitem[Yago \latin{et~al.}(2003)Yago, Kobori, Akiyama, and
  Tero-Kubota]{BQ-expt1}
Yago,~T.; Kobori,~Y.; Akiyama,~K.; Tero-Kubota,~S. Time-resolved EPR study on
  reorganization energies for charge recombination reactions in the systems
  involving hydrogen bonding. \emph{Chem. Phys. Lett.} \textbf{2003},
  \emph{369}, 49--54\relax
\mciteBstWouldAddEndPuncttrue
\mciteSetBstMidEndSepPunct{\mcitedefaultmidpunct}
{\mcitedefaultendpunct}{\mcitedefaultseppunct}\relax
\EndOfBibitem
\bibitem[Bauscher and Maentele(1992)Bauscher, and Maentele]{BQ-expt2}
Bauscher,~M.; Maentele,~W. Electrochemical and infrared-spectroscopic
  characterization of redox reactions of p-quinones. \emph{J. Phys. Chem.}
  \textbf{1992}, \emph{96}, 11101--11108\relax
\mciteBstWouldAddEndPuncttrue
\mciteSetBstMidEndSepPunct{\mcitedefaultmidpunct}
{\mcitedefaultendpunct}{\mcitedefaultseppunct}\relax
\EndOfBibitem
\bibitem[Moin \latin{et~al.}(2010)Moin, Hofer, Pribil, Randolf, and
  Rode]{Fe3_IC}
Moin,~S.~T.; Hofer,~T.~S.; Pribil,~A.~B.; Randolf,~B.~R.; Rode,~B.~M. A Quantum
  Mechanical Charge Field Molecular Dynamics Study of Fe2+ and Fe3+ Ions in
  Aqueous Solutions. \emph{Inorg. Chem.} \textbf{2010}, \emph{49},
  5101--5106\relax
\mciteBstWouldAddEndPuncttrue
\mciteSetBstMidEndSepPunct{\mcitedefaultmidpunct}
{\mcitedefaultendpunct}{\mcitedefaultseppunct}\relax
\EndOfBibitem
\bibitem[Amira \latin{et~al.}(2005)Amira, Spångberg, Zelin, Probst, and
  Hermansson]{Fe3_JPCB}
Amira,~S.; Spångberg,~D.; Zelin,~V.; Probst,~M.; Hermansson,~K.
  Car$-$Parrinello Molecular Dynamics Simulation of Fe$^{3+}$(aq). \emph{J.
  Phys. Chem. B} \textbf{2005}, \emph{109}, 14235--14242\relax
\mciteBstWouldAddEndPuncttrue
\mciteSetBstMidEndSepPunct{\mcitedefaultmidpunct}
{\mcitedefaultendpunct}{\mcitedefaultseppunct}\relax
\EndOfBibitem
\bibitem[Pasquarello \latin{et~al.}(2001)Pasquarello, Petri, Salmon, Parisel,
  Car, Éva Tóth, Powell, Fischer, Helm, and Merbach]{Cu_solv_Sc}
Pasquarello,~A.; Petri,~I.; Salmon,~P.~S.; Parisel,~O.; Car,~R.; Éva Tóth,;
  Powell,~D.~H.; Fischer,~H.~E.; Helm,~L.; Merbach,~A.~E. First Solvation Shell
  of the Cu(II) Aqua Ion: Evidence for Fivefold Coordination. \emph{Science}
  \textbf{2001}, \emph{291}, 856--859\relax
\mciteBstWouldAddEndPuncttrue
\mciteSetBstMidEndSepPunct{\mcitedefaultmidpunct}
{\mcitedefaultendpunct}{\mcitedefaultseppunct}\relax
\EndOfBibitem
\bibitem[O'Regan and Grätzel(1991)O'Regan, and Grätzel]{exp_tio2_1_nat}
O'Regan,~B.; Grätzel,~M. A low-cost, high-efficiency solar cell based on
  dye-sensitized colloidal ${\mathrm{TiO}}_{2}$ films. \emph{Nature}
  \textbf{1991}, \emph{353}, 737--740\relax
\mciteBstWouldAddEndPuncttrue
\mciteSetBstMidEndSepPunct{\mcitedefaultmidpunct}
{\mcitedefaultendpunct}{\mcitedefaultseppunct}\relax
\EndOfBibitem
\bibitem[Matthey \latin{et~al.}(2007)Matthey, Wang, Wendt, Matthiesen, Schaub,
  Lægsgaard, Hammer, and Besenbacher]{exp_tio2_2_sc}
Matthey,~D.; Wang,~J.~G.; Wendt,~S.; Matthiesen,~J.; Schaub,~R.;
  Lægsgaard,~E.; Hammer,~B.; Besenbacher,~F. Enhanced Bonding of Gold
  Nanoparticles on Oxidized ${\mathrm{TiO}}_{2}(110)$. \emph{Science}
  \textbf{2007}, \emph{315}, 1692--1696\relax
\mciteBstWouldAddEndPuncttrue
\mciteSetBstMidEndSepPunct{\mcitedefaultmidpunct}
{\mcitedefaultendpunct}{\mcitedefaultseppunct}\relax
\EndOfBibitem
\bibitem[Kowalski \latin{et~al.}(2010)Kowalski, Camellone, Nair, Meyer, and
  Marx]{Marx_PRL_dyn_tio2}
Kowalski,~P.~M.; Camellone,~M.~F.; Nair,~N.~N.; Meyer,~B.; Marx,~D. Charge
  Localization Dynamics Induced by Oxygen Vacancies on the
  ${\mathrm{TiO}}_{2}(110)$ Surface. \emph{Phys. Rev. Lett.} \textbf{2010},
  \emph{105}, 146405\relax
\mciteBstWouldAddEndPuncttrue
\mciteSetBstMidEndSepPunct{\mcitedefaultmidpunct}
{\mcitedefaultendpunct}{\mcitedefaultseppunct}\relax
\EndOfBibitem
\bibitem[Tilocca and Selloni(2004)Tilocca, and Selloni]{Selloni_meoh_ads_tio2}
Tilocca,~A.; Selloni,~A. Methanol Adsorption and Reactivity on Clean and
  Hydroxylated Anatase$(101)$ Surfaces. \emph{J. Phys. Chem. B} \textbf{2004},
  \emph{108}, 19314--19319\relax
\mciteBstWouldAddEndPuncttrue
\mciteSetBstMidEndSepPunct{\mcitedefaultmidpunct}
{\mcitedefaultendpunct}{\mcitedefaultseppunct}\relax
\EndOfBibitem
\bibitem[Di~Valentin \latin{et~al.}(2006)Di~Valentin, Pacchioni, and
  Selloni]{wat_ads_tio2_selloni}
Di~Valentin,~C.; Pacchioni,~G.; Selloni,~A. Electronic Structure of Defect
  States in Hydroxylated and Reduced Rutile ${\mathrm{TiO}}_{2}(110)$ Surfaces.
  \emph{Phys. Rev. Lett.} \textbf{2006}, \emph{97}, 166803\relax
\mciteBstWouldAddEndPuncttrue
\mciteSetBstMidEndSepPunct{\mcitedefaultmidpunct}
{\mcitedefaultendpunct}{\mcitedefaultseppunct}\relax
\EndOfBibitem
\bibitem[Morgan and Watson(2009)Morgan, and Watson]{tio2_hubbard_diff_surf}
Morgan,~B.~J.; Watson,~G.~W. A Density Functional Theory + U Study of Oxygen
  Vacancy Formation at the $(110)$, $(100)$, $(101)$, and $(001)$ Surfaces of
  Rutile ${\mathrm{TiO}}_{2}$. \emph{J. Phys. Chem. C} \textbf{2009},
  \emph{113}, 7322--7328\relax
\mciteBstWouldAddEndPuncttrue
\mciteSetBstMidEndSepPunct{\mcitedefaultmidpunct}
{\mcitedefaultendpunct}{\mcitedefaultseppunct}\relax
\EndOfBibitem
\bibitem[Calzado \latin{et~al.}(2008)Calzado, Hern\'andez, and
  Sanz]{tio2_band_gap_hubbard}
Calzado,~C.~J.; Hern\'andez,~N.~C.; Sanz,~J.~F. Effect of on-site Coulomb
  repulsion term $U$ on the band-gap states of the reduced rutile $(110)$
  $\mathrm{Ti}{\mathrm{O}}_{2}$ surface. \emph{Phys. Rev. B} \textbf{2008},
  \emph{77}, 045118\relax
\mciteBstWouldAddEndPuncttrue
\mciteSetBstMidEndSepPunct{\mcitedefaultmidpunct}
{\mcitedefaultendpunct}{\mcitedefaultseppunct}\relax
\EndOfBibitem
\bibitem[Tripathi and Nair(2012)Tripathi, and Nair]{Ravi_2012_JPCB}
Tripathi,~R.; Nair,~N.~N. Thermodynamic and Kinetic Stabilities of Active Site
  Protonation States of Class--C $\beta$-Lactamase. \emph{J. Phys. Chem. B}
  \textbf{2012}, \emph{116}, 4741--4753\relax
\mciteBstWouldAddEndPuncttrue
\mciteSetBstMidEndSepPunct{\mcitedefaultmidpunct}
{\mcitedefaultendpunct}{\mcitedefaultseppunct}\relax
\EndOfBibitem
\bibitem[Tripathi and Nair(2013)Tripathi, and Nair]{Ravi_2013_JACS}
Tripathi,~R.; Nair,~N.~N. Mechanism of Acyl--Enzyme Complex Formation from the
  {H}enry--{M}ichaelis Complex of Class--C $\beta$-Lactamases with
  $\beta$--Lactam Antibiotics. \emph{J. Am. Chem. Soc.} \textbf{2013},
  \emph{135}, 14679--14690\relax
\mciteBstWouldAddEndPuncttrue
\mciteSetBstMidEndSepPunct{\mcitedefaultmidpunct}
{\mcitedefaultendpunct}{\mcitedefaultseppunct}\relax
\EndOfBibitem
\bibitem[Das and Nair(2017)Das, and Nair]{Chandan:pccp:2017}
Das,~C.~K.; Nair,~N.~N. Hydrolysis of cephalexin and meropenem by {N}ew {D}elhi
  metallo--$\beta$--lactamase: the substrate protonation mechanism is drug
  dependent. \emph{Phys. Chem. Chem. Phys.} \textbf{2017}, \emph{19},
  13111--13121\relax
\mciteBstWouldAddEndPuncttrue
\mciteSetBstMidEndSepPunct{\mcitedefaultmidpunct}
{\mcitedefaultendpunct}{\mcitedefaultseppunct}\relax
\EndOfBibitem
\bibitem[Das and Nair(2020)Das, and Nair]{Chandan:ChemEurJ:2020}
Das,~C.~K.; Nair,~N.~N. Elucidating the Molecular Basis of Avibactam Mediated
  Inhibition of Class A $\beta$–Lactamases. \emph{Chem. Eur. J.}
  \textbf{2020}, \emph{26}, 9639--9651\relax
\mciteBstWouldAddEndPuncttrue
\mciteSetBstMidEndSepPunct{\mcitedefaultmidpunct}
{\mcitedefaultendpunct}{\mcitedefaultseppunct}\relax
\EndOfBibitem
\bibitem[Awasthi \latin{et~al.}(2018)Awasthi, Gupta, Tripathi, and
  Nair]{Shalini_AZT}
Awasthi,~S.; Gupta,~S.; Tripathi,~R.; Nair,~N.~N. Mechanism and Kinetics of
  Aztreonam Hydrolysis Catalyzed by Class--C $\beta$--Lactamase: A
  Temperature-Accelerated Sliced Sampling Study. \emph{J. Phys. Chem. B}
  \textbf{2018}, \emph{122}, 4299--4308\relax
\mciteBstWouldAddEndPuncttrue
\mciteSetBstMidEndSepPunct{\mcitedefaultmidpunct}
{\mcitedefaultendpunct}{\mcitedefaultseppunct}\relax
\EndOfBibitem
\bibitem[Slebocka-Tilk \latin{et~al.}(2002)Slebocka-Tilk, Sauriol, Monette, and
  Brown]{Formamide_Exp}
Slebocka-Tilk,~H.; Sauriol,~F.; Monette,~M.; Brown,~R.~S. Aspects of the
  hydrolysis of formamide: revisitation of the water reaction and determination
  of the solvent deuterium kinetic isotope effect in base. \emph{Can. J. Chem.}
  \textbf{2002}, \emph{80}, 1343--1350\relax
\mciteBstWouldAddEndPuncttrue
\mciteSetBstMidEndSepPunct{\mcitedefaultmidpunct}
{\mcitedefaultendpunct}{\mcitedefaultseppunct}\relax
\EndOfBibitem
\bibitem[Blumberger \latin{et~al.}(2006)Blumberger, Ensing, and
  Klein]{Angew_Klein}
Blumberger,~J.; Ensing,~B.; Klein,~M.~L. Formamide Hydrolysis in Alkaline
  Aqueous Solution: Insight from Ab Initio Metadynamics Calculations.
  \emph{Angew. Chem. Int. Ed.} \textbf{2006}, \emph{45}, 2893--2897\relax
\mciteBstWouldAddEndPuncttrue
\mciteSetBstMidEndSepPunct{\mcitedefaultmidpunct}
{\mcitedefaultendpunct}{\mcitedefaultseppunct}\relax
\EndOfBibitem
\bibitem[Blumberger and Klein(2006)Blumberger, and Klein]{CPL_Klein}
Blumberger,~J.; Klein,~M.~L. Revisiting the free energy profile for the
  nucleophilic attack of hydroxide on formamide in aqueous solution.
  \emph{Chem. Phys. Lett.} \textbf{2006}, \emph{422}, 210 -- 217\relax
\mciteBstWouldAddEndPuncttrue
\mciteSetBstMidEndSepPunct{\mcitedefaultmidpunct}
{\mcitedefaultendpunct}{\mcitedefaultseppunct}\relax
\EndOfBibitem
\end{mcitethebibliography}

\end{document}